\documentclass[12pt]{article}

\usepackage{amsmath}
\usepackage{amssymb}
\usepackage{a4wide}
\usepackage{amsmath,amsthm,amssymb,xspace}
\newtheorem{Theorem}{Theorem}[section]
\newtheorem{Lemma}{Lemma}[section]
\newtheorem{Remark}{Remark}[section]
\newtheorem{Proposition}{Proposition}[section]
\newtheorem{Definition}{Definition}[section]

\newcommand{\oneder}[2]{\ensuremath{{\mathop{{%
            \Rightarrow}}\limits^{{#1}}}_{\!\!_{#2}}\!}}
\newcommand{\multider}[1]{\ensuremath{{\mathop{{ %
            \Rightarrow}}\limits^{{#1}}}_{\!_\Re} %
            \raisebox{2pt}{\!\!\!\scriptsize *}}}
\newcommand{\multidernorm}[2]{\ensuremath{{\mathop{{%
          \Rightarrow}}\limits^{{#1}}}%
       \raisebox{2pt}{\!\scriptsize *}}\!_{\!_{#2}}\,}
\newcommand{\prsoneder}[1]{\ensuremath{\oneder{#1}{\Re}}}

\newcommand{\onederparKomega}[1]{\ensuremath{\oneder{#1}{\Re^{K,K^{\omega}}_{PAR}}}}
\newcommand{\onederseqK}[1]{\ensuremath{\oneder{#1}{\Re^{K}_{SEQ}}}}

\newcommand{\multiderparK}[1]{\ensuremath{\multidernorm{#1}{\Re^{K}_{PAR}}}}
\newcommand{\multiderparKomega}[1]{\ensuremath{\multidernorm{#1}{\Re^{K,K^{\omega}}_{PAR}}}}
\newcommand{\multiderseqK}[1]{\ensuremath{\multidernorm{#1}{\Re^{K}_{SEQ}}}}
\newcommand{\Rule}[1]{\ensuremath{{\mathop{{%
            \rightarrow}}\limits^{{#1}}}}}

\def\prsder#1#2#3{{#1\,\multidernorm{#2}{\Re}#3}}
\def\prsderpar#1#2#3{{#1\,\multidernorm{#2}{\Re_{PAR}^{K}}#3}}

\def\prsrule#1#2#3{{#1\mathop{{\rightarrow}}\limits^{{#2}}#3}}
\def\prslongrule#1#2#3{{#1\mathop{{\rightarrow}}\limits^{{#2}}#3}}

\def\prsdernorm#1#2#3#4{{\,#1\,\multidernorm{#2}{#3}#4}}

\newcommand{\maximal}[1]{$\Upsilon^{f}_{M}({#1})$}
\newcommand{\maximalG}[2]{$\Upsilon^{f}_{#1}({#2})$}
\newcommand{\maximalGInf}[2]{$\Upsilon^{\infty}_{#1}({#2})$}
\newcommand{\maximalparK}[1]{$\Upsilon^{f}_{M^{K}_{PAR}}({#1})$}
\newcommand{\maximalparKomega}[1]{$\Upsilon^{f}_{M^{K,K^{\omega}}_{PAR}}({#1})$}
\newcommand{\maximalparKomegaPlus}[1]{$\Upsilon^{f}_{M^{K,K^{\omega}}_{PAR,\infty}}({#1})$}
\newcommand{\maximalparKomegaInf}[1]{$\Upsilon^{\infty}_{M^{K,K^{\omega}}_{PAR}}({#1})$}
\newcommand{\maximalseqK}[1]{$\Upsilon^{f}_{M^{K}_{SEQ}}({#1})$}
\newcommand{\maximalseqInfK}[1]{$\Upsilon^{\infty}_{M_{SEQ}^{K}}({#1})$}
\newcommand{\maximalInf}[1]{$\Upsilon^{\infty}_{M}({#1})$}

\def\parcomp#1#2{{#1\!\parallel\!#2}}

\def\PRS{\text{{\em PRS}}\xspace}
\def\PRSs{\text{{\em PRS\/}s}\xspace}
\def\BRS{\text{{\em BRS}}\xspace}

\def\ALTL{\text{{\em ALTL}}\xspace}

\def\MBRS{\text{{\em MBRS}}\xspace}
\def\MBRSs{\text{{\em MBRS\/}s}\xspace}

\def\npla#1{{\langle #1\rangle}}

\def\ldsrule#1#2#3{\ensuremath{
   \cfrac[c]{#1}
    {\ #2\ }\text{\footnotesize \ \ensuremath{#3}}}}

\def\firstact#1{\ensuremath{\text{\emph{firstact}}}(#1)}

\title{Model checking  for Process Rewrite Systems and a class of action--based regular properties}
\author{
L.~Bozzelli$^1$ \\[10pt]
{\small\begin{tabular}{c@{\hspace{0cm}}c}
$\!\!^1$Dipartimento di Matematica e Applicazioni \\
Universit\`a di Napoli ``Federico II'' \\
Via Cintia, 80126, Napoli - Italy \\
{\tt laura.bozzelli@dma.unina.it}
\end{tabular}
}}

\date{}

\begin{document}
\maketitle
\begin{abstract}
  We consider the model checking problem
  for Process Rewrite Systems (\PRSs), an infinite-state formalism (non Turing-powerful)
  which subsumes many common models such as Pushdown Processes and
  Petri Nets. \PRSs can be adopted as formal models for programs with
  dynamic creation and synchronization of concurrent processes, and with recursive
  procedures.   The model-checking problem for
  \PRSs and action-based linear temporal logic (\ALTL) is
  undecidable. However, decidability for
  some interesting fragment of \ALTL remains an open question.
   In this paper  we state decidability results concerning
   generalized acceptance properties about infinite derivations (infinite term
  rewriting) in \PRSs. As a consequence, we obtain
  decidability of the model-checking (restricted to infinite runs)
  for \PRSs  and a meaningful fragment of \ALTL.\newline\newline
\end{abstract}

\section{Introduction}
Automatic verification of systems is nowadays one of the most
investigated topics. A major difficulty to face when considering
this problem is that reasoning about systems in general may
require dealing with infinite state models. Software systems may
introduce infinite states both manipulating data ranging over
infinite domains, and having unbounded control structures such as
recursive procedure calls and/or dynamic creation of concurrent
processes (e.g. multi--treading). Many different formalisms have
been proposed for the description of infinite state systems. Among
the most popular are the well known formalisms of Context Free
Processes, Pushdown Processes, Petri Nets, and Process Algebras.
The first two are models of sequential computation, whereas Petri
Nets and Process Algebra explicitly take into account concurrency.
The model checking problem for these infinite state formalisms
have been studied in the literature.  As far as Context Free
Processes and Pushdown Processes are concerned, decidability of
the modal $\mu$--calculus, the most powerful of the modal and
temporal logics used for verification, has been established (see
\cite{bouajjani97,burkart01,ehrs00,hun94,walukiewicz96}). In
\cite{BS94,esparza94,esp97}, model checking for Petri nets has
been studied. The branching temporal logic as well as the
state-based linear temporal logic are undecidable even for
restricted logics. Fortunately, the model checking for
action-based linear temporal logic (\ALTL)
\cite{esparza94,esp97,mayr98} is decidable.\newline Verification
of formalisms which accommodate both parallelism and recursion is
a challenging problem. In order to formally study this kind of
systems, recently the formal framework of Process Rewrite Systems
(\PRSs) has been introduced \cite{mayr98}.  This framework (non
Turing-powerful), which is based on term rewriting, subsumes many
common infinite states models such us Pushdown Processes and Petri
Nets. \PRSs can be adopted as formal models for programs with
  dynamic creation and (a restricted form of) synchronization of concurrent processes,
  and with recursive procedures.
 The decidability results already
known in the literature for the general framework of \PRSs concern
reachability analysis \cite{mayr98} and symbolic reachability
analysis \cite{bouajjani03,bouajjani04}. Unfortunately, the model
checking of action-based linear temporal logic becomes undecidable
\cite{bouajjani96,mayr98}. It remains un\-de\-ci\-dable even for
restricted models such as PA processes
 \cite{bouajjani96}. However, decidability for
  some interesting fragment of \ALTL and the
  general framework of \PRSs remains an open question.\newline

\noindent\emph{Our contribution}: In this paper we state a
decidability result concerning \emph{generalized} acceptance
properties about infinite derivations (infinite term rewriting) in
\PRSs. In order to formalize these properties we introduce the
notion of \emph{Multi B\"{u}chi Rewrite Systems} (\MBRS) that is,
informally speaking, a \PRS with a finite number of accepting
components, where each component is a subset of the \PRS.
Moreover, as a consequence of our decidability result, we obtain
decidability of the model checking (restricted to infinite runs)
for \PRSs and a meaningful fragment of \ALTL. Within this fragment
we can express important classes of properties like invariant, as
well as strong and weak fairness constraints. \newline\newline
\emph{Plan of the paper}: In Section 2, we recall the framework of
Process Rewrite Systems and \ALTL logic. In Section 3, we
introduce the notion of \emph{Multi B\"{u}chi Rewrite System}, and
show how our decidability result about generalized acceptance
properties of infinite derivations in \PRSs  can be used in
model-checking for a meaningful \ALTL fragment. In Section 4, we
prove our decidability result. Several proofs are omitted for lack
of space. They can be found in the extended version of this paper.
\newline\newline \emph{Related Work}: Our decidability result
extends
one stated in \cite{Bozz03}, regarding \emph{classical} acceptance
properties (\emph{a la} B\"{u}chi) of derivations in \PRSs. In
particular, our \ALTL fragment is strictly more expressive (and
surely more interesting in the applications) than one considered
in \cite{Bozz03}.

\section{Preliminaries}

\subsection{Process Rewrite Systems}

\begin{Definition}[Process Term] Let $Var=\{X,Y,\ldots\}$ be a finite set of process
  variables. The set  of {\em process terms} $t$ over $Var$, denoted by $T$, is
  defined by the following syntax:
  \begin{displaymath}
\textrm{$t::= \varepsilon$ $|$ $X$ $|$ $t.t$ $|$
  $t$$\parallel$$t$}
 \end{displaymath}
where $X\in{}Var$, $\varepsilon$ denotes the empty term,
``\,$\parallel$'' denotes parallel composition, and ``$.$''
denotes sequential composition.
\end{Definition}

We always work with equivalences classes of process terms modulo
commutativity and associativity of ``$\parallel$'', and modulo
associativity of ``$.$''.  Moreover $\varepsilon$ will act as the
identity for both parallel and sequential
composition\footnote{When we look at terms we think of it as
right-associative. So, when we say that a term has the form
$t_{1}.t_2$, then we mean that $t_1$ is either a single variable
or a parallel composition of process terms.}.

\begin{Definition}[Process Rewrite System]
  A {\em Process Rewrite System} $\mathrm{(}$or \PRS, or {\em Rewrite System}$\mathrm{)}$ over
  a finite
  alphabet of atomic actions $\Sigma$  and the set of process variables $Var$ is a \emph{finite}
  set of rewrite rules $\Re\subseteq{}T\times\Sigma\times{}T$ of the
  form $\prsrule{t}{a}{t'}$, where $t$ $\mathrm{(}$$\neq\varepsilon$$\mathrm{)}$ and $t'$
  are terms in $T$, and $a\in\Sigma$.
\end{Definition}

A \PRS $\Re$ over $Var$ and the alphabet $\Sigma$ induces a
labelled transition system (LTS) over $T$ with a  transition
relation $\prslongrule{}{}{} \subseteq T \times \Sigma \times T$
that is the smallest relation satisfying the following inference
rules:\newline

\ldsrule{(\prsrule{t}{a}{t'})\in\Re}{\prslongrule{t}{a}{t'}}{}\hspace*{1cm}
\ldsrule{\prslongrule{t_1}{a}{t_1'}}
     {\prslongrule{\parcomp{t_1}{t}}{a}{\parcomp{t_1'}{t}}}{}
     \hspace*{1cm}
\ldsrule{\prslongrule{t_1}{a}{t_1'}}
     {\prslongrule{\parcomp{t}{t_1}}{a}{\parcomp{t}{t_1'}}}{}
     \hspace*{1cm}
\ldsrule{\prslongrule{t_1}{a}{t_1'}}
     {\prslongrule{t.t_1}{a}{t.t_1'}}{}
\newline\newline
where $t,t',t_1,t_1'$ are process terms and $a\in\Sigma$.\newline
In similar way we define for every rule $r\in\Re$ the notion of
\emph{one--step derivation} by $r$ relation, denoted by
\prsoneder{r}.

 A \emph{path in} $\Re$ from
$t_0\in{}T$ is  a (finite or infinite) sequence of LTS edges of
the form
$\prsrule{t_{0}}{a_{0}}{t_{1}},\prsrule{t_{1}}{a_{0}}{t_{2}},\ldots$,
denoted by $\prsrule{t_{0}}{a_{0}}{t_{1}} \prsrule{}{a_{1}}{t_{2}}
\prsrule{}{a_{2}}{}\ldots$. A \emph{run in $\Re$} from $t_0$ is a
maximal path from $t_0$, i.e. a path from $t_0$ which is either
infinite or has the form $\prsrule{t_{0}}{a_{0}}{t_{1}}
\prsrule{}{a_{1}}{\ldots} \prsrule{}{a_{n-1}}{t_n}$ and there is
no edge $\prsrule{t_n}{a}{t} \in \prsrule{}{}{}$, for any $a\in
\Sigma$ and $t\in T$. We write $runs_{\Re}(t_0)$ (resp.,
$runs_{\Re,\infty}(t_0)$) to refer to the set of runs (resp.,
infinite runs) in $\Re$ from $t_0$, and $runs(\Re)$ to refer to
the set of all the runs in $\Re$.

A \emph{finite derivation} in $\Re$ from a term $t$ to a term $t'$
(through a finite sequence $\sigma = r_{1}r_{2}\ldots{}r_{n}$ of
rules in $\Re$), is a sequence $d$ of one--step derivations of the
form $t_0$ \prsoneder{r_1} $t_1$, $t_1$ \prsoneder{r_2}
$t_2$,$\ldots,$
 $t_{n-1}$ \prsoneder{r_n} $t_n$, with $t_0=t$
and $t_n=t'$, and it is denoted by $t_0$ \prsoneder{r_1} $t_1$
\prsoneder{r_2} $t_2\ldots$  $t_{n-1}$ \prsoneder{r_n} $t_n$. The
derivation $d$ is a \emph{$n$--step derivation} (or a
\emph{derivation of length $n$}), and for succinctness is also
denoted by $t$ \multider{\sigma} $t'$. Moreover, we say that $t'$
is \emph{reachable} in $\Re$ from the term $t$ (through derivation
$d$). If $\sigma$ is empty, we say that $d$ is a {\em null
derivation\/}.
\newline
An \emph{infinite derivation} in $\Re$ from a term $t_0$ (through
an infinite sequence $\sigma = r_{1}r_{2}\ldots $ of rules in
$\Re$), is an infinite sequence of one step derivations of the
form $t_0$ \prsoneder{r_1} $t_1$, $t_1$ \prsoneder{r_2}
$t_2,\ldots$, denoted by $t_0$ \prsoneder{r_1} $t_1$
\prsoneder{r_2} $t_2\ldots$. For succinctness such a derivation is
also denoted by $t_0$ \multider{\sigma}.

\bigskip

For technical reasons, we shall also consider \PRSs in a
syntactical restricted form called \emph{normal form}
\cite{mayr98}.
 A \PRS $\Re$ is said to be in {\em
    normal form} if every rule $r\in\Re$ has one of the following forms:
\begin{description}
\item[PAR rules:] $X_1$$\parallel$$X_2\ldots$$\parallel$$X_p$
\Rule{a} $Y_1$$\parallel$$Y_2\ldots$$\parallel$$Y_q\quad$ where
$p\in{}N\setminus\{0\}$ and $q\in{}N$. \item[SEQ rules:]
$\prsrule{X}{a}{Y.Z}$ or $\prsrule{X.Y}{a}{Z}$ or
  $\prsrule{X}{a}{Y}$ or $\prsrule{X}{a}{\varepsilon}$.
\end{description}
\noindent with $X,Y,Z,X_i,Y_j\in{}Var$. A \PRS where all the rules
are SEQ (resp., PAR) rules is called \emph{sequential} (resp.,
\emph{parallel}) \PRS.

\subsection{\ALTL (Action--based LTL)}

Given a finite set $\Sigma$ of atomic actions, the set of formulae
$\varphi$ of ALTL over $\Sigma$ is defined as follows:
\begin{displaymath}
\textrm{$\varphi::= true$ $|$ $\neg\varphi$ $|$
$\varphi\wedge\varphi$ $|$
  $\npla{a}\varphi$
  $|$ $\varphi\hspace{1pt}U\hspace{1pt}\varphi$ }
\end{displaymath}
where $a\in\Sigma$, $\npla{a}\varphi$ denotes the \emph{one--step
next} operator, and $U$ denotes the \emph{strong until} operator.
We also consider the derived operators
$F\varphi:=true\hspace{1pt}U\hspace{1pt}\varphi$
(``\emph{eventually} $\varphi$") and its dual $G\varphi:=\neg F
\neg \varphi$ (``\emph{always} $\varphi$").
\newline
 In order to
give semantics to ALTL formulae on a \PRS $\Re$, we need some
additional notation. Given a path $\pi=t_{0}$
$\mathop{\rightarrow}\limits^{a_{0}}$ $t_{1}$
$\mathop{\rightarrow}\limits^{a_{1}}$ $t_{2}$
$\mathop{\rightarrow}\limits^{a_{2}}\ldots$ in $\Re$, $\pi^{i}$
denotes the suffix of $\pi$ starting from the $i$--th term in the
sequence, i.e. the path $t_{i}$
$\mathop{\rightarrow}\limits^{a_{i}}$ $t_{i+1}$
$\mathop{\rightarrow}\limits^{a_{i+1}}\ldots$.   If the path $\pi$
is \emph{non--trivial} (i.e., the sequence contains at least two
terms) we denote the first action $a_{0}$  by $\firstact{\pi}$.

\ALTL formulae over a \PRS $\Re$ are interpreted in terms of the
set of the runs in $\Re$ satisfying the given \ALTL formula.  The
\emph{denotation of a formula $\varphi$} relative to $\Re$, in
symbols $[[\varphi]]_{\Re}$, is  defined inductively as follows:
\begin{itemize}
\item $[[true]]_{\Re}= runs(\Re)$, \item
$[[\neg\varphi]]_{\Re}=runs(\Re)\setminus[[\varphi]]_{\Re}$, \item
$[[\varphi_{1}\wedge\varphi_{2}]]_{\Re}=
  [[\varphi_{1}]]_{\Re}\,\cap\,[[\varphi_{2}]]_{\Re}$,
\item $[[\npla{a}\varphi]]_{\Re}=\{\pi\in{}runs(\Re)\ \mid\
  \firstact{\pi}=a \text{ and } \pi^{1}\in{}[[\varphi]]_{\Re}\}$,
\item $[[\varphi_{1}\hspace{1pt}U\hspace{1pt}\varphi_{2}]]_{\Re} =
    \{\pi\in{}runs(\Re)\ \mid\ \begin{array}[t]{l} \text{for some } i\geq
    0 \text{  }
 \pi^{i}\in{}[[\varphi_{2}]]_{\Re} \text{ and }\\ %
    \text{for all } j<i \text{ } \pi^{j}\in{}[[\varphi_{1}]]_{\Re}\
    \}.\end{array}$
\end{itemize}
 For any term $t\in{}T$
and \ALTL formula $\varphi$, we say that $t$ satisfies $\varphi$
(resp., satisfies $\varphi$ restricted to infinite runs) (w.r.t
$\Re$), in symbols $t \models_{\Re}\varphi$ (resp., $t
\models_{\Re,\infty}\varphi$), if
$runs_{\Re}(t)\subseteq[[\varphi]]_{\Re}$ (resp.,
$runs_{\Re,\infty}(t)\subseteq[[\varphi]]_{\Re}$).

The model-checking problem (resp., model--checking problem
restricted to infinite runs) for \ALTL and \PRSs is the problem of
deciding if, given a \PRS $\Re$, an \ALTL formula $\varphi$ and a
term $t$ of $\Re$, $t\models_{\Re}\varphi$ (resp.,
$t\models_{\Re,\infty}\varphi$). The following is a well--known
result:

\begin{Proposition}[see~\cite{bouajjani97,esparza94,mayr98}]\label{prop:par-seq-altl}
  The model--checking problem for \ALTL and  parallel $\mathrm{(}$resp., sequential$\mathrm{)}$ \PRSs, possibly
  restricted to infinite runs, is decidable.
\end{Proposition}

\section{Multi B\"{u}chi Rewrite Systems}

\begin{Definition}[Multi B\"{u}chi Rewrite System]
  A \emph{Multi B\"{u}chi Rewrite System} $\mathrm{(}$\emph{MBRS}$\mathrm{)}$
  $\mathrm{(}$with $n$ \emph{accepting components}$\mathrm{)}$ over a finite set of
  process variables $Var$ and an alphabet $\Sigma$ is a tuple
  $M=\npla{\Re,\npla{\Re_{1}^{A},\ldots,\Re_{n}^{A}}}$, where $\Re$ is a \PRS over $Var$
  and $\Sigma$, and
  for all $i=1,\ldots,n$ $\Re_{i}^{A}\subseteq\Re$. $\Re$ is called the \emph{support} of $M$.
\end{Definition}

In the definition above, if $n=1$, then $M$ is also called
\emph{B\"{u}chi Rewrite System} (\BRS) \cite{Bozz03}, and every
rule
$r\in\Re^{A}_{1}$ is called \emph{accepting rule} of $M$.\\
We say that $M$ is a \MBRS \emph{in
  normal form} (resp., \emph{sequential}  \MBRS, \emph{parallel} \MBRS)
if the underlying \PRS $\Re$ is in normal form (resp., is
sequential, is parallel).\\
For a  rule sequence $\sigma$
  in $\Re$  the
  \emph{finite maximal of}  $\sigma$ \emph{as to} $M$,  denoted by \maximal{\sigma}, is the
  set $\{i\in\{1,\ldots,n\}|$ $\sigma$ contains some occurrence of rule in $\Re_{i}^{A}\}$.
    The
  \emph{infinite maximal of}  $\sigma$ \emph{as to} $M$,  denoted by \maximalInf{\sigma}, is the
  set $\{i\in\{1,\ldots,n\}|$ $\sigma$ contains infinite  occurrences of some rule in $\Re_{i}^{A}\}$.
Given $K,K^{\omega}\subseteq\{1,\ldots,n\}$ and a derivation $t$
\multider{\sigma}, we say that $t$ \multider{\sigma}
  is a $(K,K^{\omega})-$\emph{accepting derivation}  in $M$ if  \maximal{\sigma} $=K$ and
  \maximalInf{\sigma} $=K^{\omega}$.\\
For all $n\in{}N\setminus\{0\}$ let us denote by $P_{n}$ the set
$2^{\{1,\ldots,n\}}$ $\mathrm{(}$i.e., the set of the subsets of
$\{1,\ldots,n\}$$\mathrm{)}$.

\subsection{Model-checking of \PRSs}\label{sec:Model-checkingPRS}

\noindent{}The main result of the paper concerns the decidability
of the following problem:

\begin{description}
 \item[Problem 1:] \emph{Given a \MBRS
    $M=\npla{\Re,\npla{\Re_{1}^{A},\ldots,\Re_{n}^{A}}}$ over Var and
    the alphabet $\Sigma$, given a process term $t$ and two
    sets
    $K,K^{\omega}\in{}P_{n}$, to decide if there exists a $(K,K^{\omega})$-accepting infinite
derivation in $M$ from $t$}.
  \end{description}

Without loss of generality we can assume that the input term $t$
in Problem 1 is a process variable in $Var$. In fact, if
$t\notin{}Var$, then, starting from $M$, we construct a new \MBRS
$M'$ by adding a new variable $X$ and a rule of the form $X$
\Rule{} $t$ whose finite maximal as to $M'$ is the empty
set.\newline

 Before proving the decidability of Problem 1 in Section \ref{sec:Result},
 we show how a solution to this problem can
be effectively exploited for automatic verification of some
meaningful (action-based) linear time properties of infinite runs
in \PRSs.  In particular, we consider the following \ALTL fragment

 \setcounter{equation}{0}
\begin{equation}\label{Eq:NewFragmentALTL}
  \varphi ::= F\,\psi\ |\ GF\,\psi \ |\ \neg \varphi \ |\  \varphi
  \wedge \varphi
\end{equation}
where $\psi$ denotes an \ALTL \emph{propositional}
formula\footnote{The set of  \ALTL propositional formulae $\psi$
over the set
$\Sigma$ of atomic actions is  defined as follows:\\
\hspace*{4cm}$\psi ::= <$$a$$>true$$ \ | \psi \wedge \psi \ |\
\neg \psi$ (where $a\in{}\Sigma$)}. For succinctness, we denote an
\ALTL propositional formula of the form $<$$a$$>true$ (with
$a\in\Sigma$) simply by $a$.
\newline
 Within this fragment, property patterns
frequent in system verification can be expressed. In particular,
we can express \emph{safety properties} (e.g., $G\,\psi_1$),
\emph{guarantee properties} (e.g., $F\,\psi_1$), \emph{obligation
properties} (e.g., $F\,\psi_1 \rightarrow F\,\psi_2$, or
$G\,\psi_1 \rightarrow G\,\psi_2$), \emph{response properties}
(e.g., $GF\,\psi_1$), \emph{persistence properties} (e.g.,
$FG\,\psi_1$), and finally \emph{reactivity properties} (e.g.,
$GF\,\psi_1 \rightarrow GF\,\psi_2$). Notice that important
classes of properties like invariants, as well as strong and weak
fairness constraints, can be expressed.\newline
 In order to prove  decidability of the
model--checking problem restricted to infinite runs for this
fragment of \ALTL we need some definitions. Given a propositional
formula $\psi$ over $\Sigma$, we denote by $[[\psi]]_{\Sigma}$ the
subset of $\Sigma$ inductively defined as follows
\begin{itemize}
    \item for all $a\in\Sigma\quad$  $[[a]]_{\Sigma}=\{a\}$,
\item $[[\neg\psi]]_{\Sigma}=\Sigma\setminus[[\psi]]_{\Sigma}$,
\item $[[\psi_{1}\wedge\psi_{2}]]_{\Sigma}=
  [[\psi_{1}]]_{\Sigma}\,\cap\,[[\psi_{2}]]_{\Sigma}$.
\end{itemize}
Evidently, given a \PRS $\Re$ over $\Sigma$, an \ALTL
propositional formula $\psi$ and an infinite run $\pi$ of $\Re$,
we have that $\pi\in[[\psi]]_{\Re}$ iff
$firstact(\pi)\in[[\psi]]_{\Sigma}$. Given a rule
$r=t$\Rule{a}$t'\in\Re$, we say that $r$ \emph{satisfies}  $\psi$
if $a\in[[\psi]]_{\Sigma}$. We denote by $AC_{\Re}(\psi)$  the set
of
 rules in $\Re$ that satisfy $\psi$. \newline

 \noindent{}Now, we can prove the following result

\begin{Theorem}\label{Theorem:NewResultOnALTL}
\noindent{}The model--checking problem for  \PRSs and the fragment
\ALTL $\mathrm{(}$\ref{Eq:NewFragmentALTL}$\mathrm{)}$, restricted
to infinite runs, is decidable.
\end{Theorem}

\begin{proof}
Given a \PRS $\Re$, a process term  $t$ and a formula $\varphi$
belonging to \ALTL fragment (\ref{Eq:NewFragmentALTL}), we have to
decide  if $t \models_{\Re,\infty} \varphi$ or, equivalently, if
there exists an infinite run $\pi\in{}runs_{\Re,\infty}(t)$
satisfying the formula
    $\neg\varphi$.\newline
Let us  consider the derived operator
$F^{+}\varphi:=F\varphi\wedge\neg{}GF\varphi$. Pushing negation
inward, and using the following logic equivalences
\begin{itemize}
    \item $G\varphi_{1}\wedge{}G\varphi_{2}\equiv{}G(\varphi_{1}\wedge\varphi_{2})$
    \item $\neg{}F\varphi_{1}\equiv{}G\neg\varphi_{1}$
    \item $\neg{}G\varphi_{1}\equiv{}F\neg\varphi_{1}$
    \item $F\varphi_{1}\equiv{}F^{+}\varphi_{1}\vee{}GF\varphi_{1}$
    \item
    $FG\varphi_{1}\equiv{}F^{+}\neg\varphi_{1}\vee{}G\varphi_{1}$
\end{itemize}
formula $\neg\varphi$ can be written in the following disjunctive
normal form
\begin{equation}\label{Eq:2}
 \neg\varphi \equiv \bigvee_{i} \Bigr( \bigwedge_{j} F^{+}\psi_{j} \wedge
 \bigwedge_{k}  GF\eta_{k} \wedge G\zeta \Bigr)
\end{equation}
where $\psi_{j}$, $\eta_{k}$, and $\zeta$ are \ALTL propositional
formulae. Evidently, we can restrict ourselves to consider a
single disjunct in (\ref{Eq:2}). In other words, our starting
problem  is reducible to the problem of deciding, given a formula
having the following form
\begin{equation}\label{Eq:3}
   F^{+}\psi_{1} \wedge\ldots\wedge F^{+}\psi_{m_{1}}\wedge
   GF\eta_{1} \wedge\ldots\wedge GF\eta_{m_{2}} \wedge
   G\zeta\footnote{$\psi_{j}$, $\eta_{k}$ and $\zeta$ are \ALTL propositional
formulae}
\end{equation}
if there exists an infinite run $\pi\in{}runs_{\Re,\infty}(t)$
satisfying formula (\ref{Eq:3}). \newline Let us consider the
\MBRS in normal form
$M=\npla{\Re,\npla{\Re_{1}^{A},\ldots,\Re_{n}^{A}}}$ where
$n=m_{1}+m_{2}+1$ and
\begin{eqnarray}
\textrm{for all $i=1,\ldots,m_{1}\quad\quad$
$\Re_{i}^{A}=AC_{\Re}(\psi_{i})$}\nonumber\\
 \textrm{for all ${}j=1,\ldots,m_{2}\quad\quad$
 $\Re_{j+m_{1}}^{A}=AC_{\Re}(\eta_{j})$}\nonumber\\
 \textrm{ $\Re_{m_{1}+m_{2}+1}^{A}=AC_{\Re}(\neg\zeta)$}\nonumber
\end{eqnarray}
Let $K=\{1,\ldots,m_{1}+m_{2}\}$ and
$K^{\omega}=\{m_{1}+1,\ldots,m_{1}+m_{2}\}$. It is easy to show
that there exists a run $\pi\in{}runs_{\Re,\infty}(t)$ satisfying
formula (\ref{Eq:3}) iff there exists a $(K,K^{\omega})$-accepting
infinite derivation in $M$ from $t$.
 By the decidability of Problem 1, we obtain the assertion.
\end{proof}

\section{Decidability results on  \MBRSs}\label{sec:Result}
In this section we prove the main result of the paper, i.e. the
decidability of  Problem \textbf{1}  defined in
Subsection~\ref{sec:Model-checkingPRS}. We proceed in two steps.
First, in Subsection~\ref{sec:infinite} we decide the problem for
the class of \MBRSs in normal form. Then,  in
Subsection~\ref{sec:infiniteUnrestricted}
 we extend the result to
the whole class of \MBRSs. For the proof we need some preliminary
results, represented by the following Propositions
\ref{Prop:ParMBRS-F}--\ref{Prop:SeqMBRS}, that easily follow from
the decidability of \ALTL model-checking problem for parallel
(resp., sequential) \PRSs (see Proposition
\ref{prop:par-seq-altl}).

\begin{Proposition}\label{Prop:ParMBRS-F}
 Given a \emph{parallel} \MBRS
  $M_{P}=\npla{\Re_{P},\npla{\Re_{P,1}^{A},\ldots,\Re_{P,n}^{A}}}$ over $Var$,
  given  two variables $X,Y\in Var$ and $K\in{}P_{n}$,
   it is decidable whether there exists a finite derivation  in $\Re_{P}$
   starting from $X$ $\mathrm{(}$resp., of the form
$\prsdernorm{X}{\sigma}{\Re_{P}}{Y}$, of the form
$\prsdernorm{X}{\sigma}{\Re_{P}}{\varepsilon}$, of the form
$\prsdernorm{X}{\sigma}{\Re_{P}}{\parcomp{t}{Y}}$ with
$|\sigma|>0$$\mathrm{)}$ such that \maximalG{M_{P}}{\sigma} $=K$.
\end{Proposition}

\begin{Proposition}\label{Prop:ParMBRS-Inf} Let us consider two \emph{parallel}
 \MBRSs
  $M_{P_{1}}=\npla{\Re_{P},\npla{\Re_{P_{1},1}^{A},\ldots,\Re_{P_{1},n}^{A}}}$ and
  $M_{P_{2}}=\npla{\Re_{P},\npla{\Re_{P_{2},1}^{A},\ldots,\Re_{P_{2},n}^{A}}}$ over $Var$,
  and with the same support $\Re_{P}$.  Given  a variable $X\in Var$,
  two sets $K,K^{\omega}\in{}P_{n}$, and a subset $\Re^{*}_{P}$ of $\Re_{P}$
  it is decidable whether there exists a  derivation in $\Re_{P}$ of
    the form $\prsdernorm{X}{\sigma}{\Re_{P}}{}$ such that \maximalG{M_{P_{1}}}{\sigma}
    $=K$, \maximalGInf{M_{P_{1}}}{\sigma} $\cup$
    \maximalG{M_{P_{2}}}{\sigma} $=K^{\omega}$, and  $\sigma$ is either
    infinite or  contains some occurrence of rule in
    $\Re_{P}\setminus\Re^{*}_{P}$.
\end{Proposition}

\noindent Now, let us give an additional notion of reachability
(for variables) in sequential \emph{PRSs}.
\begin{Definition} Given a \emph{sequential} \PRS $\Re_{S}$ over $Var$, and $X,Y\in{}Var$,
$Y$ is \emph{reachable from} $X$ \emph{in} $\Re_{S}$ if there
exists a term $t$ of the form $X_{1}.X_{2}.\ldots{}X_{n}.Y$ such
that $\prsdernorm{X}{}{\Re_{S}}{t}$.
\end{Definition}

\begin{Proposition}\label{Prop:SeqMBRS} Let us consider a \emph{sequential} \MBRS
  $M_{S}=\npla{\Re_{S},\npla{\Re_{S,1}^{A},\ldots,\Re_{S,n}^{A}}}$ over $Var$.
   Given  two variables $X,Y\in Var$ and two sets
   $K,K^{\omega}\in{}P_{n}$, it is decidable whether
\begin{enumerate}
\item $Y$ is reachable from $X$ in $\Re_{S}$ through a derivation
having finite maximal $K$ as to $M_{S}$.
 \item  There exists a
$(K,K^{\omega})$-accepting infinite derivation in  $M_{S}$ from
$X$.
\end{enumerate}
\end{Proposition}

\subsection{Decidability of Problem 1 for \MBRSs in normal form}\label{sec:infinite}

\setcounter{equation}{0} In this subsection we prove the
decidability of Problem 1 restricted to the class of \MBRSs in
normal form. We shall use the following result stated
in~\cite{Bozz03}.

\begin{Theorem}[see~\cite{Bozz03}]\label{Theorem:OldResults}
\noindent Given a \BRS $M=\npla{\Re,\Re_F}$ in normal form and a
        process variable $X$ it is decidable whether
        there exists an infinite derivation in
       $\Re$ from $X$ of the form $X$ \multider{\sigma} such that
       $\sigma$ does \emph{not} contain  occurrences of accepting
       rules.
\end{Theorem}

Let $M=\npla{\Re,\npla{\Re_{1}^{A},\ldots,\Re_{n}^{A}}}$ be a
\MBRS \emph{in normal form} over $Var$ and the alphabet $\Sigma$,
and $K$ and $K^{\omega}$ be elements in $P_{n}$. Given
$X\in{}Var$, we have to decide if there exists a
$(K,K^{\omega})$-accepting infinite derivation in $M$ from $X$.
The proof of decidability is by induction on $|K|+|K^{\omega}|$.\\
 \textbf{Base Step}: $|K|=0$ and $|K^{\omega}|=0$. Let $M_{F}=\npla{\Re,\Re_{F}}$
 be the \BRS with $\Re_F=\bigcup_{i=1}^{n}\Re_{i}^{A}$. Given an
 infinite derivation $X$ \multider{\sigma} in $\Re$ from a variable $X$, then
this derivation is $(\emptyset,\emptyset)$-accepting in $M$ if,
and only if, it does not contain occurrences of accepting rules in
$M_{F}$. So, the decidability result follows from Theorem
\ref{Theorem:OldResults}.\\
  \textbf{Inductive Step}:  $|K|+|K^{\omega}|>0$. By the
inductive hypothesis, for each $K'\subseteq{}K$ and
$K'^{\omega}\subseteq{}K^{\omega}$ with
 $|K'|+|K'^{\omega}|<|K|+|K^{\omega}|$ the
result holds. Starting from this assumption we shall show that
Problem 1, with input the sets $K$ and $K^{\omega}$, can be
reduced to (a combination of) two similar, but simpler, problems
(that are decidable): the first (resp., the second) is a
decidability problem on infinite derivations of parallel (resp.,
sequential) \MBRSs. Before illustrating our approach, we need few
additional definitions and notation.

\begin{Remark} \emph{Since $M$ is in normal form we can limit ourselves to
consider only  terms $t$, called \emph{terms in normal form},
defined as $t ::= X \mid t$$\parallel$$t \mid X.t$ $($where
$X\in{}Var$$)$. In fact, given a term in normal form $t$, each
term $t'$ reachable from $t$ in $M$ is still in normal form.}
\end{Remark}

 In the following,
$M_{P}=\npla{\Re_{P},\npla{\Re_{P,1}^{A},\ldots,\Re_{P,n}^{A}}}$
denotes the restriction of  $M$  to the PAR rules, i.e. $\Re_{P}$
(resp., $\Re_{P,i}^{A}$ for $i=1,\ldots,n$) is the set $\Re$
(resp., $\Re_{i}^{A}$ for $i=1,\ldots,n$) restricted to the PAR
rules. Moreover, we shall use two new variables  $\hat{Z}_{F}$ and
$\hat{Z}_{\infty}$, and denote by $T$ (resp., $T_{PAR}$,
$T_{SEQ}$)  the set of process terms in normal form (resp.,  in
which no sequential composition occurs,  in which no  parallel
composition occurs) over
$Var\cup\{\hat{Z}_{F},\hat{Z}_{\infty}\}$.

\begin{Definition}[Subderivation]
  Let $\overline{t}$ \multider {\lambda}
  $t$$\parallel$$(X.s)$ \multider{\sigma} be a
  derivation in $\Re$  from $\overline{t}\in{}T$. The set of the
  \emph{subderivations $d'$ of
    $d=(t$$\parallel$$(X.s)$ \multider{\sigma}$)$ from $s$} is inductively
  defined as follows:

\begin{enumerate}
\item if $d$ is a null derivation \emph{or} $s = \varepsilon$
\emph{or} $d$ is of the form $t$$\parallel$$(X.Z)$ \prsoneder{r}
$t$$\parallel$$Y$ \multider{\sigma'}  $\mathrm{(}$with $r
=X.Z$\Rule{a}$Y$ and $s=Z\in{}Var$$\mathrm{)}$,
     then $d'$ is
  the null derivation from $s$;
 \item if $d$ is of the form
     $t$$\parallel$$(X.s)$ \prsoneder{r} $t$$\parallel$$(X.s')$
    \multider{\sigma'}   $\mathrm{(}$with  $s$ \prsoneder{r} $s'$$\mathrm{)}$
and $s'$ \multider{\mu'} is a subderivation of
$t$$\parallel$$(X.s')$ \multider{\sigma'} from $s'$, then $s$
\prsoneder{r} $s'$ \multider{\mu'} is a subderivation of $d$ from
$s$;
 \item if $d$ is of the form
     $t$$\parallel$$(X.s)$ \prsoneder{r} $t'$$\parallel$$(X.s)$
    \multider{\sigma'}   $\mathrm{(}$with  $t$ \prsoneder{r} $t'$$\mathrm{)}$,
    then every  subderivation of $t'$$\parallel$$(X.s)$
\multider{\sigma'} from $s$ is also a subderivation of $d$ from
$s$.
\end{enumerate}
Moreover, we say that $d'$ is a subderivation of $\overline{t}$
\multider{\lambda}
  $t$$\parallel$$(X.s)$ \multider{\sigma}.
\end{Definition}

Given a rule sequence $\sigma$  in $\Re$, and a subsequence
$\sigma'$ of $\sigma$, $\sigma\setminus \sigma{}'$ denotes the
rule sequence obtained by removing from $\sigma$ all and only the
occurrences of rules in $\sigma'$.\newline Let us denote by
$\Pi^{K,K^{\omega}}_{PAR,\infty}$ the set of derivations $d$
 in $\Re$ such that there does \emph{not} exist  a subderivation of $d$
    that is  a $(K,K^{\omega})$-accepting infinite derivation in
    $M$.

\noindent{}Let us sketch the main idea of our technique. At first,
let us focus  on the class of derivations
$\Pi^{K,K^{\omega}}_{PAR,\infty}$. Let $p$ \multider{\sigma} be a
$(\overline{K},\overline{K}^{\omega})$-accepting derivation in $M$
belonging to $\Pi^{K,K^{\omega}}_{PAR,\infty}$ with
$p\in{}T_{PAR}$, $\overline{K}\subseteq{}K$ and
$\overline{K}^{\omega}\subseteq{}K^{\omega}$. The idea is to mimic
this derivation  by using only PAR rules belonging to extensions
of the parallel \MBRS $M_{P}$.
   If $\sigma$
contains only PAR rule occurrences, then $p$ \multider{\sigma} is
a $(\overline{K},\overline{K}^{\omega})$-accepting derivation  in
the parallel \MBRS $M_{P}$. Otherwise, $p$ \multider{\sigma} can
be written in the form:
 \begin{equation}\label{eq:derivation}
 \textrm{$p$ \multider{\lambda} $\overline{p}$$\parallel$$X$ \prsoneder{r}
 $\overline{p}$$\parallel$$(Y.Z)$ \multider{\omega}}
 \end{equation}
 where $r = X$\Rule{a}$Y.Z$,  $\lambda$ contains only occurrences of PAR rules
 in $\Re$,  $\overline{p}\in{}T_{PAR}$ and $X,Y,Z\in{}Var$. Let $Z$ \multider{\rho} be a subderivation of
 $\overline{p}$$\parallel$$(Y.Z)$ \multider{\omega} from $Z$. By the
 definition of subderivation  only one of the following four cases may occur:
\begin{description}
\item[A]  $Z$ \multider{\rho} is finite and  $\overline{p}$
\multider{\omega\setminus\rho}. \item[B] $Z$ \multider{\rho} leads
to the term $\varepsilon$, and
  $p$ \multider{\sigma} is of the form
$p$ \multider{\lambda} $\overline{p}$$\parallel$$X$ \prsoneder{r}
 $\overline{p}$$\parallel$$(Y.Z)$ \multider{\omega_{1}} $t$$\parallel$$Y$
 \multider{\omega_{2}},
  where $\rho$ is a subsequence of $\omega_1$ and $\overline{p}$
  \multider{\omega_{1}\setminus\rho} $t$.
\item[C] $Z$ \multider{\rho} leads to a variable $W\in{}Var$, and
  $p$ \multider{\sigma} can be written as
\begin{equation}\label{eq:form1}
 \textrm{$p$ \multider{\lambda} $\overline{p}$$\parallel$$X$ \prsoneder{r}
 $\overline{p}$$\parallel$$(Y.Z)$ \multider{\omega_{1}} $t$$\parallel$$(Y.W)$
 \prsoneder{r'} $t$$\parallel$$W'$ \multider{\omega_{2}}}
\end{equation}
  where $r'=Y.W$\Rule{b}$W'$ (with $W'\in{}Var$),
   $\rho$ is a subsequence of $\omega_1$ and $\overline{p}$
  \multider{\omega_{1}\setminus\rho} $t$.
\item[D]  $Z$ \multider{\rho} is infinite, and $\overline{p}$
\multider{\omega\setminus\rho}.
 \end{description}
Cases \textbf{A}, \textbf{B} and \textbf{C} are similar, so for
brevity we examine only cases \textbf{C} and \textbf{D}. At first,
let us consider case \textbf{C}.
  The derivation in equation (\ref{eq:form1})  is
  $(\overline{K},\overline{K}^{\omega})$-accepting if, and only if, the following
  derivation, obtained by anticipating the application of the
  rules in $\rho$ before the application of the rules in
  $\xi=\omega_{1}\setminus\rho$,
  is $(\overline{K},\overline{K}^{\omega})$-accepting
  \begin{equation}\label{eq:form1-2}
   \textrm{$p$ \multider{\lambda} $\overline{p}$$\parallel$$X$ \prsoneder{r}
 $\overline{p}$$\parallel$$(Y.Z)$ \multider{\rho} $\overline{p}$$\parallel$$(Y.W)$  \prsoneder{r'} $\overline{p}$$\parallel$$W'$
    \multider{\xi}
 $t$$\parallel$$W'$ \multider{\omega_{2}}}
  \end{equation}
The idea is to collapse the finite derivation $X$ \prsoneder{r}
$Y.Z$
 \multider{\rho} $Y.W$ \prsoneder{r'} $W'$ into a single PAR rule
 of the form $r''=X$\Rule{K'}$W'$ where $K'=$ \maximal{rr'\rho} $\subseteq{}K$.
 So, the label of $r''$ keeps track of the finite maximal of
 $rr'\rho$ in $M$.
  Now, we can apply recursively the same reasoning to the
   derivation in $\Re$ from $\overline{p}$$\parallel$$W'\in{}T_{PAR}$
    given by $\overline{p}$$\parallel$$W'$ \multider{\xi} $t$$\parallel$$W'$
      \multider{\omega_{2}},
  which belongs to $\Pi^{K,K^{\omega}}_{PAR,\infty}$ and whose finite (resp., infinite)
  maximal as to $M$ is
  contained in $K$ (resp., $K^{\omega}$). Now, let us consider
  case \textbf{D}.
Since $p$ \multider{\sigma} belongs to
$\Pi^{K,K^{\omega}}_{PAR,\infty}$, we have that
  \maximal{\rho} $=K_1\subseteq{}K$, \maximalInf{\rho} $=K_{1}^{\omega}\subseteq{}K^{\omega}$,
\maximal{r} $=K_2\subseteq{}K$, \maximalInf{r} $=\emptyset$
  and  $|K_1|+|K_{1}^{\omega}|<|K|+|K^{\omega}|$.
  From our assumptions (inductive hypothesis)
   it is decidable whether  there exists a
  $(K_1,K_{1}^{\omega})$-accepting infinite derivation in $M$ from
  variable $Z$.
  Then, we keep track of  the infinite rule sequence $r\rho$
  by adding
  a  PAR rule of the form $r'=X$\Rule{K',K_{1}^{\omega}}$\hat{Z}_{\infty}$  with
  $K'=K_1\cup{}K_2$. So, the label of $r'$ keeps track of the finite and infinite maximal of
 $r\rho$ in $M$.
   Now, we can apply recursively the same reasoning to the
   derivation $\overline{p}$$\parallel$$\hat{Z}_{\infty}$ \multider{\omega\setminus\rho} in $\Re$ from
   $\overline{p}$$\parallel$$\hat{Z}_{\infty}\in{}T_{PAR}$,
  which belongs to $\Pi^{K,K^{\omega}}_{PAR,\infty}$ and whose finite (resp., infinite)
  maximal as to $M$ is
  contained in $K$ (resp., $K^{\omega}$).

In other words,  all  subderivations in $p$ \multider{\sigma} are
abstracted away by PAR rules not belonging to $\Re$,
 according to the intuitions given above.\newline
 For keeping track of the finite subderivations of the forms \textbf{A},
 \textbf{B} and \textbf{C}, we define a first extension of the parallel \MBRS $M_P$ in the following
 way.

\begin{Definition}\label{def:MPAR-K}
 The \MBRS $M^{K}_{PAR}=\npla{\Re^{K}_{PAR},\npla{\Re^{K,A}_{PAR,1},\ldots,\Re_{PAR,n}^{K,A}}}$
  is
 the least parallel \MBRS with $n$ accepting components, over $Var\cup\{\hat{Z}_{F}\}$ and the alphabet $\overline{\Sigma} = \Sigma
  \cup{}P_{n}$\footnote{let us assume that $\Sigma\cap{}P_{n}=\emptyset$}, satisfying the following properties:
\begin{enumerate}
    \item $\Re^{K}_{PAR}\supseteq\Re_{P}$
    and
    $\Re^{K,A}_{PAR,i}\supseteq\Re^{A}_{P,i}$
    for all $i=1,\ldots,n$.
    \item  Let $r=X$\Rule{a}$Y.Z\in{}\Re$,
    $Z$ \multiderparK{\sigma} $p$ for some term $p$ $\mathrm{(}$resp., $Z$ \multiderparK{\sigma} $\varepsilon$$\mathrm{)}$,
    and
     $K'=$ \maximal{r} $\cup$ \maximalparK{\sigma}.
     If $K'\subseteq{}K$, then $r'=X$\Rule{K'}$\hat{Z}_{F}\in{}\Re^{K}_{PAR}$
     $\mathrm{(}$resp., $r'=X$\Rule{K'}$Y\in{}\Re^{K}_{PAR}$$\mathrm{)}$ and
            \maximalparK{r'} $=K'$.
    \item Let $r=X$\Rule{a}$Y.Z\in{}\Re$,
    $r'=Y.W$\Rule{b}$W'\in{}\Re$,
     $Z$ \multiderparK{\sigma} $W$, and  $K'=$ \maximal{rr'} $\cup$ \maximalparK{\sigma}.
            If $K'\subseteq{}K$, then
             $r''=X$\Rule{K'}$W'\in{}\Re^{K}_{PAR}$ and
            \maximalparK{r''} $=K'$.
    \end{enumerate}
\end{Definition}

\setcounter{equation}{0}
\begin{Lemma}\label{Lemma:Algo}
The parallel \MBRS
$M^{K}_{PAR}=\npla{\Re^{K}_{PAR},\npla{\Re^{K,A}_{PAR,1},\ldots,\Re_{PAR,n}^{K,A}}}$
 can be effectively constructed.
\end{Lemma}
\begin{proof}
Figure \ref{fig:algo} reports the procedure
BUILD-PARALLEL-MBRS($M$,$K$), which, starting from the \MBRS $M$
(in normal form) and the set $K\in{}P_{n}$, builds the parallel
\MBRS
$M^{K}_{PAR}=\npla{\Re^{K}_{PAR},\npla{\Re^{K,A}_{PAR,1},\ldots,\Re_{PAR,n}^{K,A}}}$.
The algorithm uses the routine \emph{UPDATE}$(r',K')$ that is
defined as follows:\newline\newline
$\Re_{PAR}^{K}:=\Re_{PAR}^{K}\cup\{r'\};$\\
\textbf{for each} $i\in{}K'$ \textbf{do}
$\Re_{PAR,i}^{A,K}:=\Re_{PAR,i}^{A,K}\cup\{r'\};$\newline

 Notice that by
Proposition \ref{Prop:ParMBRS-F}, the conditions in each of the
\textbf{if} statements in lines 7, 9 and 13 are decidable,
therefore, the procedure is effective. Moreover, since the set of
rules of the form $X$\Rule{K'}$Y$  with $X\in{}Var$,
$Y\in{}Var\cup\{\hat{Z}_{F}\}$ and $K'\in{}P_{n}$ is finite,
termination immediately follows.

\begin{figure}[htbp]
\ \\\noindent \textbf{Algorithm}
  B{\small{}UILD}--{\small{}PARALLEL}--MBRS($M$,$K$)\vspace{4pt}
\\\small
1 $\Re_{PAR}^{K}:=\Re_{P}$;\\
2 \textbf{for} $i=1,\ldots,n$ \textbf{do} $\Re_{PAR,i}^{A,K}:=\Re_{P,i}^{A}$;\\
3 \textbf{repeat}\\
4 $\quad\quad{}$\emph{flag:=false;}\\
5 $\quad\quad{}$\textbf{for each} $r=X$\Rule{a}$Y.Z\in\Re$ \emph{and} $K_{1}\subseteq{}K$ \emph{such that} \maximal{r} $\subseteq{}K$  \textbf{do}\\
6 $\quad\quad\quad\quad$\emph{Set}
            $K'=K_{1}$ $\cup$ \maximal{r}; \\
7 $\quad\quad\quad\quad$\textbf{if}
$\prsdernorm{Z}{\sigma}{\Re_{PAR}^{K}}{p}$ \emph{for some} $p$
            \emph{such that} \maximalG{M_{PAR}^{K}}{\sigma} $=K_{1}$ \textbf{then}\\
8 $\quad\quad\quad\quad\quad\quad{}$\textbf{if}
        $r'=X$\Rule{K'}$\hat{Z}_{F}\notin{}\Re_{PAR}^{K}$ \textbf{then}
        \emph{UPDATE}$(r',K')$; \emph{flag:=true;}\\
9 $\quad\quad\quad\quad$\textbf{if}
$\prsdernorm{Z}{\sigma}{\Re_{PAR}^{K}}{\varepsilon}$
            \emph{such that} \maximalG{M_{PAR}^{K}}{\sigma} $=K_{1}$ \textbf{then}\\
10 $\quad\quad\quad\quad\quad\quad{}$\textbf{if}
        $r'=X$\Rule{K'}$Y\notin{}\Re_{PAR}^{K}$ \textbf{then} \emph{UPDATE}$(r',K')$; \emph{flag:=true;}\\
11 $\quad\quad\quad\quad$\textbf{for each} $r'=Y.W$\Rule{b}$W'\in\Re$  \emph{such that} \maximal{r'} $\subseteq{}K$ \textbf{do}\\
12 $\quad\quad\quad\quad\quad\quad{}$\emph{Set}
            $K'=K_{1}$ $\cup$ \maximal{rr'}; \\
13 $\quad\quad\quad\quad\quad\quad{}$\textbf{if}
$\prsdernorm{Z}{\sigma}{\Re_{PAR}^{K}}{W}$
            \emph{such that} \maximalG{M_{PAR}^{K}}{\sigma} $=K_{1}$ \textbf{then}\\
14 $\quad\quad\quad\quad\quad\quad\quad\quad{}$\textbf{if}
        $r''=X$\Rule{K'}$W'\notin{}\Re_{PAR}^{K}$ \textbf{then} \emph{UPDATE}$(r'',K')$; \emph{flag:=true;} \\
15 \textbf{until} \emph{flag = false} \\
\caption{Algorithm to build the parallel  \MBRS
$M^{K}_{PAR}$.}\label{fig:algo}
\end{figure}
\end{proof}

\noindent In  order to simulate infinite subderivations of the
form \textbf{D}, we need to add additional PAR rules in
$M_{PAR}^{K}$.
 The following definition provides an extension of
$M_{PAR}^{K}$ suitable for our purposes.

\begin{Definition}\label{Def:M-PARs-Infinite}
By
    $M^{K,K^{\omega}}_{PAR}=
   \npla{\Re^{K,K^{\omega}}_{PAR},\npla{\Re^{K,K^{\omega},A}_{PAR,1},\ldots,
   \Re^{K,K^{\omega},A}_{PAR,n}}}$
   and \newline
   $M^{K,K^{\omega}}_{PAR,\infty}=
   \npla{\Re^{K,K^{\omega}}_{PAR},\npla{\Re^{K,K^{\omega},A}_{PAR,\infty,1},\ldots,
   \Re^{K,K^{\omega},A}_{PAR,\infty,n}}}$ we denote the parallel \MBRSs
   over $Var\cup\{\hat{Z}_{F},\hat{Z}_{\infty}\}$ and the alphabet
   $\Sigma\cup{}P_n\cup{}P_n\times{}P_n$ $($with the same support$)$,
  defined by $M$ and $M_{PAR}^{K}$
   in the following way:
\begin{itemize}
\item $\Re^{K,K^{\omega}}_{PAR}=\begin{array}[t]{l}%
    \Re_{PAR}^{K}\text{ }\cup \\
    \{\prsrule{X}{\overline{K},\overline{K}^{\omega}}{\hat{Z}_{\infty}} \mid \begin{array}[t]{l}
         \overline{K}\subseteq{}K,
         \overline{K}^{\omega}\subseteq{}K^{\omega},
          \text{ there exists a
         rule } r=\prsrule{X}{a}{Y.Z}\in\Re \\[4pt]
       \text{and
             an infinite  derivation } \prsder{Z}{\sigma}{} \text{
         such that } \\[4pt] |\text{\maximal{\sigma}}| +
         |\text{\maximalInf{\sigma}}|
         < |K|+|K^{\omega}| \text{ and} \\[4pt]
        \text{ \maximal{\sigma} $\cup$ \maximal{r} } =\overline{K} \text{ and}
           \text{ \maximalInf{\sigma} } =\overline{K}^{\omega}
         \}\
     \end{array}
  \end{array}$
\item
   $\Re^{K,K^{\omega},A}_{PAR,i}=\Re_{PAR,i}^{K,A}\cup\{
  X$\Rule{\overline{K},\overline{K}^{\omega}}$\hat{Z}_{\infty}\in\Re^{K,K^{\omega}}_{PAR}|$
  $i\in{}\overline{K}$\} for all $i=1,\ldots,n$
\item
  $\Re^{K,K^{\omega},A}_{PAR,i,\infty}=\{
  X$\Rule{\overline{K},\overline{K}^{\omega}}$\hat{Z}_{\infty}\in\Re^{K,K^{\omega}}_{PAR}|$
  $i\in{}\overline{K}^{\omega}$\} for all $i=1,\ldots,n$
\end{itemize}
\end{Definition}

\noindent{}By the inductive hypothesis on  decidability of Problem
1 for sets $K',K^{'\omega}\in{}P_{n}$ such that $K'\subseteq{}K$,
$K^{'\omega}\subseteq{}K^{\omega}$ and
$|K'|+|K^{'\omega}|<|K|+|K^{\omega}|$, it follows that
\begin{Lemma}
$M^{K,K^{\omega}}_{PAR}$ and $M^{K,K^{\omega}}_{PAR,\infty}$ can
be built effectively.
\end{Lemma}

\noindent The following two lemmata establish the validity of our
construction.

\begin{Lemma}\label{Lemma:From-M-To-MPAROMEGA}
 Let $p$ \multider{\sigma}$ $ be a $(\overline{K},\overline{K}^{\omega})$-accepting
  derivation  in $M$ belonging to
 $\Pi^{K,K^{\omega}}_{PAR,\infty}$,
 with $p\in{}T_{PAR}$, $\overline{K}\subseteq{}K$ and
 $\overline{K}^{\omega}\subseteq{}K^{\omega}$.  Then,
 there exists in $\Re^{K,K^{\omega}}_{PAR}$ a derivation of the form $p$ \multiderparKomega{\rho}
 such that \maximalparKomega{\rho} = $\overline{K}$ and
 \maximalparKomegaInf{\rho} $\cup$ \maximalparKomegaPlus{\rho} =
 $\overline{K}^{\omega}$. Moreover,
 if $\sigma$ is infinite, then $\rho$ is either
    infinite or  contains some occurrence of rule in
    $\Re^{K,K^{\omega}}_{PAR}\setminus\Re_{PAR}^{K}$.
\end{Lemma}

\begin{Lemma}\label{Lemma:From-MPAR-To-M-Inf}
Let $p$ \multiderparKomega{\sigma}  with $p\in{}T_{PAR}$. Then,
there exists in $\Re$ a derivation  of the form $p$
\multider{\delta} such that \maximal{\delta} =
\maximalparKomega{\sigma} and \maximalInf{\delta} =
\maximalparKomegaInf{\sigma} $\cup$ \maximalparKomegaPlus{\sigma}.
Moreover, if $\sigma$ is either infinite or contains some
occurrence of rule in
$\Re^{K,K^{\omega}}_{PAR}\setminus\Re_{PAR}^{K}$, then $\delta$ is
infinite.
\end{Lemma}

 Now, let us go back to Problem 1 and consider a
$(K,K^{\omega})$-accepting infinite derivation in $M$ from a
variable $X$ of the form $X$ \multider{\sigma}, and non belonging
to $\Pi^{K,K^{\omega}}_{PAR,\infty}$.  In this case, the
derivation $X$ \multider{\sigma} can be written in the form $X$
\multider{} $t$$\parallel$$(Y.Z)$ \multider{\rho}, with
$Z\in{}Var$, and such that there exists a subderivation of
$t$$\parallel$$(Y.Z)$ \multider{\rho} from $Z$ that is a
$(K,K^{\omega})$-accepting infinite derivation in $M$. In order to
manage this kind of derivation, we build, starting from the \MBRSs
$M$ and $M_{PAR}^{K}$, a \emph{sequential} \MBRS $M_{SEQ}^{K}$
according to the following definition:

\begin{Definition}\label{Def:M-SEQ}
  By $M_{SEQ}^{K}=\npla{\Re_{SEQ}^{K},\npla{\Re_{SEQ,1}^{K,A},\ldots,\Re_{SEQ,n}^{K,A}}}$ we denote
  the \emph{sequential} \MBRS
  over $Var$ and the alphabet $\overline{\Sigma}=\Sigma\cup{}P_{n}$  defined as follows:
\begin{itemize}
\item $\Re_{SEQ}^{K}=\begin{array}[t]{l}%
    \{\prsrule{X}{a}{Y.Z}\in\Re\}\ \cup \\
    \{\prsrule{X}{K'}{Y} \mid \begin{array}[t]{l}
         X,Y\in{}Var, K'\subseteq{}K \text{ and there exists a
         derivation } \prsderpar{X}{\sigma}{\parcomp{p}{Y}} \\[4pt]
       \text{ in } \Re_{PAR}^{K} \text{
         for some } p\in{}T_{PAR}
    \text{, with } |\sigma|>0 \text{ and  \maximalparK{\sigma} = $K'$} \}\
     \end{array}
  \end{array}$
\item
$\Re_{SEQ,i}^{K,A}=\{X$\Rule{a}$Y.Z\in\Re_{i}^{A}\}\cup\{X$\Rule{K'}$Y\in\Re_{SEQ}^{K}\mid
i\in{}K'$\} for all $i=1,\ldots,n$
\end{itemize}
\end{Definition}

\noindent By Proposition \ref{Prop:ParMBRS-F} we obtain the
following result
\begin{Lemma}
$M_{SEQ}^{K}$ can be built effectively.
\end{Lemma}

\setcounter{equation}{0}

\noindent Soundness and completeness of the procedure described
above is stated by the following two theorems.

\begin{Theorem}\label{Theorem:Problem1.1}
Let  $K\neq{}K^{\omega}$. Given $X\in{}Var$, there exists a
 $(K,K^{\omega})$-accepting infinite derivation  in $M$ from
 $X$ if, and only if,  the following property is satisfied:
\begin{itemize}
    \item There exists a variable  $Y\in{}Var$ reachable from $X$ in
    $\Re_{SEQ}^{K}$ through a $(K',\emptyset)$-accepting derivation in $M_{SEQ}^{K}$
    with $K'\subseteq{}K$, and there exists a derivation
    $Y$ \multiderparKomega{\rho}
    such that \maximalparKomega{\rho} = $K$
    and
    \maximalparKomegaInf{\rho} $\cup$ \maximalparKomegaPlus{\rho} =
    $K^{\omega}$. Moreover,  $\rho$ is either
    infinite or  contains some occurrence of rule in
    $\Re^{K,K^{\omega}}_{PAR}\setminus\Re_{PAR}^{K}$.
\end{itemize}
\end{Theorem}

\begin{Theorem}\label{Theorem:Problem1.2}
Let  $K=K^{\omega}$. Given $X\in{}Var$, there exists a
 $(K,K^{\omega})$-accepting infinite derivation  in $M$ from
 $X$ if, and only if, one of  the following conditions is satisfied:
\begin{enumerate}
    \item There exists a variable  $Y\in{}Var$ reachable from $X$ in
    $\Re_{SEQ}^{K}$ through a $(K',\emptyset)$-accepting derivation in $M_{SEQ}^{K}$
    with $K'\subseteq{}K$, and there exists a derivation
    $Y$ \multiderparKomega{\rho}
    such that \maximalparKomega{\rho} = $K$
    and
    \maximalparKomegaInf{\rho} $\cup$ \maximalparKomegaPlus{\rho} =
    $K^{\omega}$. Moreover,  $\rho$ is either
    infinite or  contains some occurrence of rule in
    $\Re^{K,K^{\omega}}_{PAR}\setminus\Re_{PAR}^{K}$.
    \item There exists a
     $(K,K^{\omega})$-accepting infinite derivation  in $M_{SEQ}^{K}$ from
     $X$.
\end{enumerate}
\end{Theorem}

These two results, together  with Propositions
\ref{Prop:ParMBRS-Inf} and \ref{Prop:SeqMBRS}, allow us to
conclude that
 Problem 1 restricted to the class of \MBRSs in normal form is
 decidable.

\subsection{Decidability of Problem 1 for unrestricted \MBRSs}\label{sec:infiniteUnrestricted}

In this section  we extend the decidability result stated in the
previous Subsection to the whole class of \MBRSs, showing that
Problem \textbf{1} for unrestricted \MBRSs is reducible to the
Problem \textbf{1} for \MBRSs in normal form. We use a
construction very close to one used in \cite{mayr98} to solve the
reachability problem for \PRSs. Remember that we can assume that
the input term in Problem \textbf{1} is a process
variable.\newline
 Let $M$ be a
\MBRS over $Var$ and the alphabet $\Sigma$, and with $n$ accepting
components. Now, we describe a procedure that transforms $M$ into
a new \MBRS $M'$ with the same number of accepting components.
Moreover, this procedure has in input also a finite set of rules
$\Re_{AUX}$, and transforms it in $\Re'_{AUX}$. If $M$ is not in
normal form, then there exists a rule in $M$ that is neither a PAR
rule nor a SEQ rule. We call such rules \emph{bad rules}
\cite{mayr98}. There are five types of bad rules\footnote{Remember
that we assume that sequential composition is right-associative.
So, when we write $t_{1}.t_2$, then  $t_1$ is either a single
variable or a parallel composition of process terms.}:

\begin{enumerate}
    \item The bad rule is $r=u$\Rule{a}$u_1$$\parallel$$u_2$. Let
    $Z_1,Z_2,W$ be new variables (non belonging to $Var$). We get $M'$ replacing the bad rule
    $r$ with the rules $r'=u$\Rule{}$W$, $r_3=W$\Rule{}$Z_1$$\parallel$$Z_2$,
    $r_1=Z_1$\Rule{}$u_1$, $r_2=Z_2$\Rule{}$u_2$ such that
    \maximalG{M'}{r'} = \maximalG{M}{r}, \maximalG{M'}{r_1}
    = \maximalG{M'}{r_2} = \maximalG{M'}{r_3} =
    $\emptyset$. If $r\in\Re_{AUX}$, then
    $\Re'_{AUX}=(\Re_{AUX}\setminus\{r\})\cup\{r',r_1,r_2,r_3\}$.
    Otherwise, $\Re'_{AUX}=\Re_{AUX}$.
    \item The bad rule is $r=u_1$$\parallel$$(u_2.u_3)$\Rule{a}$u$. Let
    $Z_1,Z_2$ be new variables. We get $M'$ replacing the bad rule
    $r$ with the rules
    $r_1=u_1$\Rule{}$Z_1$, $r_2=u_2.u_3$\Rule{}$Z_2$, $r'=Z_1$$\parallel$$Z_2$\Rule{a}$u$ such that
    \maximalG{M'}{r'} = \maximalG{M}{r}, \maximalG{M'}{r_1} = \maximalG{M'}{r_2} =
    $\emptyset$. If $r\in\Re_{AUX}$, then
    $\Re'_{AUX}=(\Re_{AUX}\setminus\{r\})\cup\{r',r_1,r_2\}$.
    Otherwise, $\Re'_{AUX}=\Re_{AUX}$.
    \item  The bad rule is $r=u$\Rule{a}$u_1.u_2$ (resp., $r=u_1.u_2$\Rule{a}$u$)
    where $u_1$ is not a single variable. Let
    $Z$ be a new variable. We get $M'$ and $\Re'_{AUX}$ in two steps. First, we
    substitute $Z$ for $u_1$ in (left-hand and right-hand sides of) all the
    rules of $M$ and $\Re_{AUX}$. Then, we add the rules $r_1=Z$\Rule{}$u_1$ and
    $r_2=u_1$\Rule{}$Z$ such that \maximalG{M'}{r_1} = \maximalG{M'}{r_2} =
    $\emptyset$.
    \item The bad rule is $r=u_1$\Rule{a}$X.u_2$ where $u_2$ is not a single variable. Let
    $Z,W$ be  new variables. We get $M'$ replacing the bad rule
    $r$ with the rules $r'=u_1$\Rule{}$W$, $r_1=W$\Rule{}$X.Z$,
    $r_2=Z$\Rule{}$u_2$ such that
    \maximalG{M'}{r'} = \maximalG{M}{r} and \maximalG{M'}{r_1} = \maximalG{M'}{r_2} =
    $\emptyset$. If $r\in\Re_{AUX}$, then
    $\Re'_{AUX}=(\Re_{AUX}\setminus\{r\})\cup\{r',r_1,r_2\}$.
    Otherwise, $\Re'_{AUX}=\Re_{AUX}$.
    \item  The bad rule is $r=X.u_1$\Rule{a}$u_2$ where $u_1$ is not a single variable. Let
    $Z$ be a new variable. We get $M'$ replacing the bad rule
    $r$ with the rules $r_1=u_1$\Rule{}$Z$, $r'=X.Z$\Rule{}$u_2$,
     such that
    \maximalG{M'}{r'} = \maximalG{M}{r} and \maximalG{M'}{r_1} =
    $\emptyset$. If $r\in\Re_{AUX}$, then
    $\Re'_{AUX}=(\Re_{AUX}\setminus\{r\})\cup\{r',r_1\}$.
    Otherwise, $\Re'_{AUX}=\Re_{AUX}$.
\end{enumerate}

 After a finite number of applications of this
procedure, starting from $\Re_{AUX}=\emptyset$, we obtain a \MBRS
$M'$ in normal form\footnote{ Note that we have not specified the
label of the new rules, since it is not relevant.} and a finite
set of rules $\Re'_{AUX}$.
 Let $M'=\npla{\Re',\npla{\Re^{'A}_{1},\ldots,\Re_{n}^{'A}}}$.
Now, let us consider the \MBRS in normal form with $n+1$ accepting
components given by
$M_F=\npla{\Re',\npla{\Re_{1}^{'A},\ldots,\Re_{n}^{'A},\Re'\setminus\Re'_{AUX}}}$.
 We can prove that, given a variable
$X\in{}Var$ and two sets $K,K^{\omega}\in{}P_n$, there exists a
$(K,K^{\omega})$-accepting infinite derivation in $M$ from $X$ if,
and only if, there exists a
$(K\cup\{n+1\},K^{\omega}\cup\{n+1\})$-accepting infinite
derivation in $M_F$ from $X$.
\newline

 \noindent{\Large \textbf{Conclusion}}\newline

 In this paper we have stated decidability about generalized
acceptance properties of infinite derivations in \PRSs. Our result
has an immediate application to the model--checking within a
meaningful fragment of \ALTL logic. In order to obtain this result
we have used an approach different from classical
automata--theoretic one. The reason is that \PRSs are not closed
under intersection with state finite ($\omega$-star-free)
automaton \cite{bouajjani96} (and in fact model-checking for full
\ALTL is undecidable). Future work should aim to extend our result
to a larger fragment of \ALTL. In particular, we are working on
the \ALTL fragment (closed under boolean operations) which uses
the temporal operators $G$ (``\emph{always}'') and  $F$
(``\emph{eventually}'') without restrictions (i.e. nested
arbitrarily).

\bibliographystyle{plain}


\newpage
\appendix
\begin{LARGE}
\textbf{APPENDIX}
\end{LARGE}

\section{Definitions and simple properties}


In this section we give some definitions and deduce simple
properties that will be used in  sections  B--C for the proof of
Lemmata
\ref{Lemma:From-M-To-MPAROMEGA}--\ref{Lemma:From-MPAR-To-M-Inf}
and  Theorems \ref{Theorem:Problem1.1}--\ref{Theorem:Problem1.2}.

In the following $\hat{Var}$ denotes the set of variables
$Var\cup\{\hat{Z}_{F},\hat{Z}_{\infty}\}$, $T$ denotes the set of
terms \emph{in normal form} over $\hat{Var}$, and $T_{PAR}$
(resp., $T_{SEQ}$) the set of terms in $T$ not containing
sequential (resp., parallel) composition.

\begin{Definition}
The set of {\em subterms\/} of a term $t\in{}T$, denoted by
$SubTerms(t)$, is defined inductively as follows:
\begin{itemize}
\item $SubTerms(\varepsilon)=\{\varepsilon\}$. \item
$SubTerms(X)=\{X\}$, for all  $X\in{}\hat{Var}$. \item
$SubTerms(X.t)=SubTerms(t)\cup\{X.t\}$, for all
      $X\in{}\hat{Var}$ and $t\in{}T\setminus \{\varepsilon\}$. \item
$SubTerms($$t_{1}$$\parallel$$t_{2})=\bigcup_{(t_{1}',t_{2}')\in{}S}
(SubTerms(t_{1}')\cup{} SubTerms(t_{2}'))$ $\cup$
$\{t_{1}$$\parallel$$t_{2}\}$,
\newline
with $S=\{(t_{1}',t_{2}')\in{}T\times{}T \mid
t_{1}',t_{2}'\neq\varepsilon$
     and $t_{1}$$\parallel$$t_{2}={}t_{1}'$$\parallel$$t_{2}'\}$ and
$t_{1},t_{2}\in{}T\setminus \{\varepsilon\}$\footnote{Remember
that we identify terms with their equivalence classes. In
particular, $t_1 = t_2$ (resp., $t_1 \not= t_2$) is used to mean
that $t_1$ is equivalent (resp., not equivalent) to $t_2$. }.
\end{itemize}
\end{Definition}

\begin{Definition} The set of terms obtained from a term $t\in T$
{\em substituting\/} an occurrence of a subterm $st$ of $t$ with a
term $t'\in T$, denoted by $t[st\rightarrow{}t']$, is defined
inductively as follows:
\begin{itemize}
    \item $t[t\rightarrow{}t']=\{t'\}$.
    \item $X.t[st\rightarrow{}t']=
     \{X.s \mid s\in{}t[st\rightarrow{}t']\}$, for all
      $X\in{}\hat{Var}$, $t\in{}T\setminus \{\varepsilon\}$ and
     $st\in{}SubTerms(X.t)\setminus \{X.t\}$.
    \item $t_{1}$$\parallel$$t_{2}[st\rightarrow{}t']=$
$\{\parcomp{t''}{t'_2} \mid (t'_1,t'_2) \in T \times T, t'_1,t'_2
\neq \varepsilon$, $\parcomp{t'_1}{t'_2} = \parcomp{t_1}{t_2}$,
$st \in SubTerms(t'_1)$, $t''\in{}t_{1}'[st\rightarrow{}t']\}$,
for all $t_1,t_2\in T\setminus\{\varepsilon\}$ and $st \in
SubTerms(\parcomp{t_1}{t_2})\setminus \{\parcomp{t_1}{t_2}\}$.
\end{itemize}
\end{Definition}

\begin{Definition}
  For a term $t\in{}T$, the set of terms $SEQ(t)$ is the subset of
  $T_{SEQ}\setminus\{\varepsilon\}$ defined inductively as follows:
\begin{itemize}
    \item $SEQ(\varepsilon)=\emptyset$.
    \item $SEQ(X)=\{X\}$, for all $X\in{}\hat{Var}$.
    \item $SEQ(X.t)=\{X.t' \mid t'\in{}SEQ(t)\}$, for all
      $X\in{}\hat{Var}$ and $t\in{}T\setminus \{\varepsilon\}$.
\item $SEQ(t_{1}$$\parallel$$t_{2})=SEQ(t_{1})\cup{}SEQ(t_{2})$.
    \end{itemize}
\end{Definition}

For a term $t\in{}T_{SEQ}\setminus \{\varepsilon\}$ having the
form $t=X_{1}.X_{2}.\ldots{}X_{n}.Y$, we denote by $last(t)$ the
variable $Y$. Given two terms
 $t,t'\in{}T_{SEQ}\setminus\{\varepsilon\}$, with
 $t=X_{1}.X_{2}.\ldots{}X_{n}.Y$ and
 $t'=X_{1}'.X_{2}'.\ldots{}X_{k}'.Y'$, we denote by
 $t\circ{}t'$ the term
 $X_{1}.X_{2}.\ldots{}X_{n}.X_{1}'.X_{2}'$
 $.\ldots{}X_{k}'.Y'$. Notice that $t\circ{}t'$  is
the only term  in $t[Y\rightarrow{}t']$, and that the operation
$\circ$ on terms in $T_{SEQ} \setminus \{\varepsilon\}$ is
associative.

The proof of the following two Propositions is simple

\begin{Proposition}
\label{Prop:Subterms1} The following properties hold:
\begin{enumerate}
    \item If $t$ \multider{\sigma} $t'$ and $t\in{}SubTerms(s)$,
for some $s\in{}T$, then it holds $s$ \multider{\sigma} $s'$ for
all $s'\in{}s[t\rightarrow{}t']$; \item If $t$ \multider{\sigma}
is an infinite derivation in $\Re$ and $t\in{}SubTerms(s)$, for
some $s\in{}T$, then it holds $s$ \multider{\sigma}.
\end{enumerate}
\end{Proposition}


\begin{Proposition}\label{Prop:Subterms2}
Let $\Re_S$ be a sequential \PRS over $Var$. If
 $t,t'\in{}T_{SEQ}\setminus \{\varepsilon\}$ such that
$last(t)$ $\multidernorm{\rho}{\Re_{S}}$
 $t'$, then it holds that
 \begin{enumerate}
    \item $t$ $\multidernorm{\rho}{\Re_{S}}$ $t\circ{}t'$;
    \item $t''\circ{}t$ $\multidernorm{\rho}{\Re_{S}}$ $t''\circ{}t\circ{}t'$
 for all $t''\in{}T_{SEQ}\setminus\{\varepsilon\}$.
 \end{enumerate}
\end{Proposition}



\noindent{}Now, we give the notion of \emph{Interleaving} of a
(finite or infinite) sequence of rule sequences in a \PRS $\Re'$.
In order to formalize this concept and facilitate the proof of
some connected results, we redefine the notion of sequence rule.
Precisely, a sequence rule in $\Re'$ can be seen as a mapping
$\sigma:N'\rightarrow{}\Re'$ where $N'$ can be a generic subset of
$N$.  A rule sequence $\sigma':N''\rightarrow{}\Re'$ is a
subsequence of $\sigma:N'\rightarrow{}\Re'$ iff $N''\subseteq{}N'$
and $\sigma'=\sigma|_{N''}$, that is $\sigma'$
 is the restriction of $\sigma$ to the set $N''$.
For a rule sequence $\sigma:N'\rightarrow{}\Re'$, we denote by
$pr(\sigma)$ the set $N'$. For a set $N'\subseteq{}N$ we denote by
$min(N')$ the smallest element of $N'$. Given two rule sequences
$\sigma$ and $\sigma'$, we say that they are disjoint if
$pr(\sigma)\cap{}pr(\sigma')=\emptyset$.\\
Let $n\in{}N\setminus\{0\}$ and   $(K_{h})_{h=0}^{m}$ be a
sequence of elements in $P_{n}$ (where $m\in{}N\cup\{\infty\}$).
Let us
 denote by $\bigoplus_{h=0}^{m}K_{h}$ the element of
 $P_{n}$ given by $\{i|$ for all $j\in{}N$ there exists a $h>j$
 such that $i\in{}K_{h}\}$. Evidently, if $m$ is finite, then
$\bigoplus_{h=0}^{m}K_{h}$ is empty.

\begin{Definition}\label{Def:Interleaving}
Let $(\rho_{h})_{h=0}^{m}$ be a  sequence of rule sequences in a
\PRS $\Re'$ $\mathrm{(}$where $m\in{}N\cup\{\infty\}$$\mathrm{)}$.
The \emph{Interleaving} of $(\rho_{h})_{h=0}^{m}$, denoted by
$Interleaving((\rho_{h})_{h=0}^{m})$, is the set of rule sequences
$\sigma$ in $\Re'$ such that there exists an \emph{injective}
mapping
$M_{\sigma}:\bigcup_{h=0}^{m}(\{h\}\times{}pr(\rho_{h}))\rightarrow{}N$
$\mathrm{(}$depending on $\sigma$$\mathrm{)}$ satisfying the
following properties $\mathrm{(}$where $\Delta$ is the set
$\bigcup_{h=0}^{m}(\{h\}\times{}pr(\rho_{h})$$\mathrm{))}$
\begin{itemize}
    \item For all $h=1,\ldots,m$ and for all $n,n'\in{}pr(\rho_{h})$ with
     $n<n'$, then $M_{\sigma}(h,n)<M_{\sigma}(h,n')$;
    \item $pr(\sigma)=M_{\sigma}(\Delta)$;
    \item for all $(h,n)\in{}\Delta\quad$
    $\sigma(M_{\sigma}(h,n))=\rho_{h}(n)$.
\end{itemize}
\end{Definition}

\noindent The proof of the following two Propositions is simple.

\begin{Proposition}\label{Prop:MaximalInterleaving}
Let
  $M'$ be
  a  \MBRS with support $\Re'$, and  $(\sigma_{h})_{h=0}^{m}$ be a sequence of rule
  sequences in $\Re'$ $\mathrm{(}$where $m\in{}N\cup\{\infty\}$$\mathrm{)}$. Then,
    for all $\pi\in{}Interleaving((\sigma_{h})_{h=0}^{m})$ we have
\begin{enumerate}
    \item \maximalG{M'}{\pi} = $\bigcup_{h=0}^{m}$\maximalG{M'}{\sigma_{h}}.
    \item \maximalGInf{M'}{\pi} =
    $\bigcup_{h=0}^{m}$\maximalGInf{M'}{\sigma_{h}} $\cup$
                           $\bigoplus_{h=0}^{m}$\maximalG{M'}{\sigma_{h}}.
\end{enumerate}
\end{Proposition}

\begin{Proposition}\label{Prop:Interleaving}
Let $\sigma$ be a rule sequence in a \PRS $\Re'$ and
$(\rho_{h})_{h=0}^{m}$  $\mathrm{(}$where
$m\in{}N\cup\{\infty\}$$\mathrm{)}$ be a sequence of subsequences
of $\sigma$ two by two disjoints and such that
$\bigcup_{h=0}^{m}pr(\rho_{h})=pr(\sigma)$. Then,
$\sigma\in{}Interleaving((\rho_{h})_{h=0}^{m})$.
\end{Proposition}

\section{Proof of Lemmata \ref{Lemma:From-M-To-MPAROMEGA} and \ref{Lemma:From-MPAR-To-M-Inf}}

\begin{Remark}\label{Remark:MPAR-Inf}By construction, the
following properties hold:
\begin{itemize}
    \item for all $r\in\Re_{PAR}^{K}\quad$   \maximalparKomega{r} =
          \maximalparK{r} and \maximalparKomegaPlus{r}
          $=\emptyset$.
    \item for all $r\in\Re^{K,K^{\omega}}_{PAR}\cap\Re\quad$   \maximalparKomega{r} =
          \maximal{r} and \maximalparKomegaPlus{r}
          $=\emptyset$.
    \item for all $r=X$\Rule{\overline{K},\overline{K}^{\omega}}$\hat{Z}_{\infty}\in
    \Re^{K,K^{\omega}}_{PAR}\quad$
   \maximalparKomega{r} = $\overline{K}$ and
          \maximalparKomegaPlus{r} = $\overline{K}^{\omega}$.
\end{itemize}
\end{Remark}


\noindent The following lemma easily follows by the definition of
subderivation.
\begin{Lemma}\label{Lemma:Subderivations1}
 Let $t$$\parallel$$(X.s)$ \multider{\sigma} be a
derivation in $\Re$, and let $s$ \multider{\sigma{}'} be a
subderivation of  $t$$\parallel$$(X.s)$ \multider{\sigma} from
$s$. Then, one of the following conditions is satisfied:
\begin{enumerate}
    \item $s$ \multider{\sigma{}'} is infinite and
    $t$ \multider{\sigma\setminus\sigma{}'}.
     Moreover, if $t$$\parallel$$(X.s)$ \multider{\sigma}
    is in $\Pi^{K,K^{\omega}}_{PAR,\infty}$,  then also  $t$ \multider{\sigma\setminus\sigma{}'} is in
    $\Pi^{K,K^{\omega}}_{PAR,\infty}$.
    \item $s$ \multider{\sigma{}'} leads
    to $\varepsilon$ and the derivation
    $t$$\parallel$$(X.s)$ \multider{\sigma} can be written in the form
    \begin{displaymath}
     \textrm{$t$$\parallel$$(X.s)$ \multider{\sigma_{1}} $t'$$\parallel$$X$ \multider{\sigma_{2}}}
    \end{displaymath}
    where  $t$ \multider{\lambda} $t'$ and $\sigma_{1}\in{}Interleaving(\lambda,\sigma{}')$.
    Moreover, if $t$$\parallel$$(X.s)$ \multider{\sigma}
    is in $\Pi^{K,K^{\omega}}_{PAR,\infty}$, there is a derivation
    of the form $t$$\parallel$$X$ \multider{\lambda} $t'$$\parallel$$X$ \multider{\sigma_{2}}
    belonging to $\Pi^{K,K^{\omega}}_{PAR,\infty}$.
    \item $s$ \multider{\sigma{}'} leads
    to a term $s'\neq\varepsilon$ and $t$ \multider{\sigma\setminus\sigma{}'}.
    If  $t$$\parallel$$(X.s)$ \multider{\sigma}
            is in $\Pi^{K,K^{\omega}}_{PAR,\infty}$, then also $t$ \multider{\sigma\setminus\sigma{}'}
            is in
            $\Pi^{K,K^{\omega}}_{PAR,\infty}$. Moreover, if
            $t$$\parallel$$(X.s)$ \multider{\sigma} is finite
            and leads to $\overline{t}$, then
            $\overline{t}=(X.s')$$\parallel$$t'$ where $t$
            \multider{\sigma\setminus\sigma{}'} $t'$.
     \item $s$ \multider{\sigma{}'} leads to a variable $W\in{}Var$
      and the derivation
        $t$$\parallel$$(X.s)$ \multider{\sigma} can be written in the
        form
        \begin{displaymath}
        \textrm{$t$$\parallel$$(X.s)$ \multider{\sigma_{1}}
        $t'$$\parallel$$(X.W)$ \prsoneder{r}
        $t'$$\parallel$$W'$ \multider{\sigma_{2}}}
        \end{displaymath}
        where  $r=X.W$\Rule{a}$W'\in{}\Re$,
         $t$ \multider{\lambda} $t'$
         and $\sigma_{1}\in{}Interleaving(\lambda,\sigma{}')$.
          Moreover, if $t$$\parallel$$(X.s)$ \multider{\sigma}
        is in $\Pi^{K,K^{\omega}}_{PAR,\infty}$, there is a derivation
        of the form $t$$\parallel$$W'$ \multider{\lambda} $t'$$\parallel$$W'$ \multider{\sigma_{2}}
    belonging to $\Pi^{K,K^{\omega}}_{PAR,\infty}$.
\end{enumerate}
\end{Lemma}

\subsection{Proof of Lemma \ref{Lemma:From-M-To-MPAROMEGA}}

In order to prove Lemma \ref{Lemma:From-M-To-MPAROMEGA}, we need
the following Lemma.
\begin{Lemma}\label{Lemma:From-M-To-MKPAR}
 Let $p$ \multider{\sigma} $t$$\parallel$$p'$
 with $p,p'\in{}T_{PAR}$ and \maximal{\sigma} $\subseteq{}K$. Then,
 there exists a $s\in{}T_{PAR}$ such that $p$ \multiderparK{\rho} $s$$\parallel$$p'$
 with
\maximal{\sigma} = \maximalparK{\rho}, and $s=\varepsilon$ if
$t=\varepsilon$.
\end{Lemma}
\begin{proof}
The proof is by induction on the length of finite derivations $p$
\multider{\sigma} in $\Re$ from terms in $T_{PAR}$ with
\maximal{\sigma} $\subseteq{}K$, and uses Lemma
\ref{Lemma:Subderivations1}, Properties 2--3 in the Definition of
$M^{K}_{PAR}$ and Remark \ref{Remark:MPAR-Inf}. For brevity, we
omit it.
\end{proof}

 \setcounter{equation}{0}
 Now, we can prove Lemma \ref{Lemma:From-M-To-MPAROMEGA}.
 Let $p$ \multider{\sigma}$ $ be a $(\overline{K},\overline{K}^{\omega})$-accepting
 non--null derivation  in $M$ belonging to
 $\Pi^{K,K^{\omega}}_{PAR,\infty}$,
 with $p\in{}T_{PAR}$, $\overline{K}\subseteq{}K$ and
 $\overline{K}^{\omega}\subseteq{}K^{\omega}$.  We have to prove
 that
 there exists in $\Re^{K,K^{\omega}}_{PAR}$ a derivation of the form $p$ \multiderparKomega{\rho}
  such that \maximalparKomega{\rho} = $\overline{K}$ and
 \maximalparKomegaInf{\rho} $\cup$ \maximalparKomegaPlus{\rho} =
 $\overline{K}^{\omega}$. Moreover,
 if $\sigma$ is infinite, then  $\rho$ is either
    infinite or  contains some occurrence of rule in
    $\Re^{K,K^{\omega}}_{PAR}\setminus\Re_{PAR}^{K}$.
    At first, let us prove the following property
\begin{description}
    \item[A] There exists a $p'\in{}T_{PAR}$, a non empty finite rule
    sequence $\lambda$ in $\Re^{K,K^{\omega}}_{PAR}$, and a non
    empty subsequence $\eta$ (possibly infinite) of $\sigma$ such that
    $min(pr(\eta))=min(pr(\sigma))$ (i.e. the first rule
        occurrence in $\eta$ is the first rule occurrence in
        $\sigma$), $p$ \multiderparKomega{\lambda} $p'$,
    \maximalparKomega{\lambda} = \maximal{\eta},
    \maximalparKomegaPlus{\lambda} = \maximalInf{\eta},
    $p'$ \multider{\sigma\setminus\eta} and this derivation
        is in $\Pi^{K,K^{\omega}}_{PAR,\infty}$.
    Moreover, if $\sigma$ is infinite, then either $\sigma\setminus\eta$
        is infinite or $\lambda$ is a rule in $\Re^{K,K^{\omega}}_{PAR}\setminus\Re_{PAR}^{K}$.
\end{description}

\noindent The derivation $p$ \multider{\sigma} can be rewritten as
\begin{equation}
\textrm{$p$ \prsoneder{r} $t$ \multider{\sigma'} }
\end{equation}
At first, let us assume that r is a PAR rule. In this case
$t\in{}T_{PAR}$ and
     $r\in\Re^{K,K^{\omega}}_{PAR}$. By Remark \ref{Remark:MPAR-Inf}
    \maximalparKomega{r} = \maximal{r}, and
    \maximalparKomegaPlus{r} = $\emptyset$ = \maximalInf{r}.
    Moreover,
    $t$ \multider{\sigma'}
        is in $\Pi^{K,K^{\omega}}_{PAR,\infty}$ with
        $\sigma'=\sigma\setminus{}r$.
    Thus, since $\sigma'$ is infinite if $\sigma$ is infinite,
    property \textbf{A} follows,  setting $p'=t$, $\lambda=r$ and
    $\eta=r$. If $r$ is not a PAR rule, then
$r=Z$\Rule{a}$Y.Z'$ (since $p\in{}T_{PAR}$) for some
$Z,Y,Z'\in{}Var$ and $a\in\Sigma$. So, $p=p''$$\parallel$$Z$ and
$t=p''$$\parallel$$(Y.Z')$ with $p''\in{}T_{PAR}$. From (1), let
$Z'$ \multider{\nu} be  a subderivation of
$t=p''$$\parallel$$(Y.Z')$ \multider{\sigma'} from $Z'$.  By Lemma
\ref{Lemma:Subderivations1} we can distinguish four subcases.
Since cases 2--4 (of Lemma \ref{Lemma:Subderivations1}) are
similar, for brevity, we consider only cases 1 and 4.
\newline\newline
\textbf{Case 1}: $Z'$ \multider{\nu} is infinite, and
$p''$ \multider{\sigma'\setminus\nu}.
        Moreover, $p''$ \multider{\sigma'\setminus\nu}
                is in $\Pi^{K,K^{\omega}}_{PAR,\infty}$.
By the hypothesis,
$($\maximal{\nu},\maximalInf{\nu}$)\neq(K,K^{\omega})$,
 \maximal{\nu} $\subseteq{}K$ and
 \maximalInf{\nu} $\subseteq{}K^{\omega}$.
 Hence,   $|$\maximal{\nu}$|+|$\maximalInf{\nu}$|<|K|+|K^{\omega}|$.
 Moreover, $r=Z$\Rule{a}$Y.Z'$ with \maximal{r} $\subseteq{}K$.
 By the definition of $\Re^{K,K^{\omega}}_{PAR}$, it follows that
  $r'=Z$\Rule{K_{1},K^{\omega}_{1}}$\hat{Z}_{\infty}\in\Re^{K,K^{\omega}}_{PAR}\setminus\Re_{PAR}^{K}$
  where $K_{1}$ = \maximal{\nu} $\cup$ \maximal{r} and
    $K^{\omega}_{1}$ = \maximalInf{\nu}.
     By Remark \ref{Remark:MPAR-Inf}, we have that
     \maximalparKomega{r'} = $K_{1}$ and \maximalparKomegaPlus{r'} = $K^{\omega}_{1}$.
     So, we have that $p=p''$$\parallel$$Z$ \onederparKomega{r'}
  $p''$$\parallel$$\hat{Z}_{\infty}$.
  Moreover, $p''$$\parallel$$\hat{Z}_{\infty}$
  \multider{\sigma'\setminus\nu} and this derivation is in $\Pi^{K,K^{\omega}}_{PAR,\infty}$.
  Since
                $\sigma'\setminus\nu=\sigma\setminus{}r\nu$ and
                \maximalparKomegaPlus{r'} = \maximalInf{\nu} = \maximalInf{r\nu},
                property \textbf{A} follows, setting
                $p'=p''$$\parallel$$\hat{Z}_{\infty}$, $\lambda=r'$ and
                $\eta=r\nu$.
  \newline\newline
\textbf{Case 4}: $Z'$ \multider{\nu} leads to a variable
$W\in{}Var$ and the derivation $p''$$\parallel$$(Y.Z')$
                \multider{\sigma'}  can be rewritten as $p''$$\parallel$$(Y.Z')$
                \multider{\sigma_{1}} $t'$$\parallel$$(Y.W)$
                \prsoneder{r'} $t'$$\parallel$$W'$
                \multider{\sigma_{2}}, with $p''$ \multider{\sigma'_{1}} $t'$,
                $r'=Y.W$\Rule{b}$W'$ and
                $\sigma_{1}\in{}Interleaving(\nu,\sigma'_{1})$.
              Moreover, $p''$$\parallel$$W'$
                \multider{\sigma'_{1}} $t'$$\parallel$$W'$
                \multider{\sigma_{2}} and this derivation  is in
                $\Pi^{K,K^{\omega}}_{PAR,\infty}$.
Since $Z'$ \multider{\nu} $W$
                and
                \maximal{\nu} $\subseteq$ $K$, by Lemma
                \ref{Lemma:From-M-To-MKPAR} it follows that $Z'$ \multiderparK{\chi}
                $W$  with
                \maximalparK{\chi} = \maximal{\nu}.
              Since $r=Z$\Rule{a}$Y.Z'\in\Re$ and
              $r'=Y.W$\Rule{a}$W'\in\Re$, where \maximal{r} $\subseteq$
              $K$ and
              \maximal{r'} $\subseteq$ $K$, by the definition of
               $\Re_{PAR}^{K}$ it follows that
              $r''=Z$\Rule{K'}$W'\in\Re_{PAR}^{K}$
                where
                $K'=$ \maximal{rr'} $\cup{}$ \maximalparK{\chi} = \maximal{r\nu{}r'}  and
                \maximalparK{r''} = $K'$. By construction,
                $r''\in\Re^{K,K^{\omega}}_{PAR}$, and
                by Remark \ref{Remark:MPAR-Inf} \maximalparKomega{r''} = $K'$ and
                \maximalparKomegaPlus{r''} = $\emptyset$.
               Since
                $\sigma\setminus{}r\nu{}r'=\sigma'_{1}\sigma_{2}$,
                \maximalparKomegaPlus{r''} = $\emptyset$ = \maximalInf{r\nu{}r'},
                and $\sigma'_{1}\sigma_{2}$ is
                infinite if $\sigma$ is infinite,
                property \textbf{A} follows setting
                $p'=p''$$\parallel$$W'$, $\lambda=r''$ and
                $\eta=r\nu{}r'$.

Therefore, Property \textbf{A} is satisfied. Since
$\sigma\setminus\eta$ is a subsequence of $\sigma$, we have
\maximal{\sigma\setminus\eta} $\subseteq{}K$ and
\maximalInf{\sigma\setminus\eta} $\subseteq{}K^{\omega}$. Thus, if
$\sigma\neq\eta$ we can apply property \textbf{A} to the
derivation $p'$ \multider{\sigma\setminus\eta}. Repeating this
reasoning it follows that there exists a $m\in{}N\cup\{\infty\}$,
a sequence $(p_{h})_{h=0}^{m+1}$ of terms in $T_{PAR}$, a sequence
$(\lambda_{h})_{h=0}^{m}$ of non empty finite rule sequences in
$\Re^{K,K^{\omega}}_{PAR}$, two sequences $(\sigma_{h})_{h=0}^{m}$
and $(\eta_{h})_{h=0}^{m}$ of non empty rule sequences in $\Re$
such that for all $h=0,\ldots,m\quad$
\begin{enumerate}
    \item[1.] $p=p_{0}$ and $\sigma=\sigma_{0}$.
    \item[2.]  $\eta_{h}$ is a subsequence of
    $\sigma_{h}$, $min(pr(\eta_{h}))=min(pr(\sigma_{h}))$, and if
    $h\neq{}m$ then $\sigma_{h+1}=\sigma_{h}\setminus\eta_{h}$.
     \item[3.]
        $p_{h}$ \multiderparKomega{\lambda_{h}} $p_{h+1}$,
    \maximalparKomega{\lambda_{h}} = \maximal{\eta_{h}},
        \maximalparKomegaPlus{\lambda_{h}} =
        \maximalInf{\eta_{h}}, and $p_{h}$
    \multider{\sigma_{h}}.
    \item[4.] If $m$ is finite, then $\sigma_{m}=\eta_{m}$.
    If $\sigma$ is infinite, then either  $m$ is infinite or
    there exists an $h$ such that $\lambda_{h}$ is a rule in
    $\Re^{K,K^{\omega}}_{PAR}\setminus\Re_{PAR}^{K}$.
\end{enumerate}
\noindent{}By setting $\rho=\lambda_{0}\lambda_{1}\ldots$ we have
that $p$ \multiderparKomega{\rho}. By Property 4 it follows that
if $\sigma$ is infinite, then either $\rho$ is infinite or $\rho$
contains some occurrence of rule in
$\Re^{K,K^{\omega}}_{PAR}\setminus\Re_{PAR}^{K}$.  Let us assume
that $m=\infty$. The proof for $m$  finite is simpler. By
Properties 1--2 $\eta_{0},\eta_{1},\ldots$ are non empty
subsequences of $\sigma$ two by two disjoints. Since $\sigma$ is
infinite,  we can assume that $pr(\sigma)=N$. Now, let us show
that
\begin{enumerate}
    \item[5.] $\sigma\in{}Interleaving((\eta_{h})_{h\in{}N})$
\end{enumerate}
By Proposition \ref{Prop:Interleaving} it suffices to prove that
for all $h\in{}N$ there exists an $i\in{}N$ such that
$h\in{}pr(\eta_{i})$. By Property 2 it follows  that for all
$h\in{}N$ $min(pr(\sigma_{h}))< min(pr(\sigma_{h+1}))$. Let
$h\in{}N$, then there exists the smallest $i\in{}N$ such that
$h\notin{}pr(\sigma_{i})$. Since $\sigma_0=\sigma$, $i>0$ and
$h\in{}pr(\sigma_{i-1})$. Since
$\sigma_{i}=\sigma_{i-1}\setminus\eta_{i-1}$,
$h\notin{}pr(\sigma_{i})$ and $h\in{}pr(\sigma_{i-1})$, it follow
that $h\in{}pr(\eta_{i-1})$. Thus, Property 5 holds. By Properties
3, 5, and Proposition \ref{Prop:MaximalInterleaving} it follows
that \maximalparKomega{\rho} $=$
$\bigcup_{h\in{}N}$\maximalparKomega{\lambda_{h}} $=$
$\bigcup_{h\in{}N}$\maximal{\eta_{h}} $=$ \maximal{\sigma} $=$
$\overline{K}$. Moreover,
 \begin{displaymath}
\textrm{\maximalparKomegaInf{\rho} $\cup$
\maximalparKomegaPlus{\rho} $=$
$\bigoplus_{h\in{}N}$\maximalparKomega{\lambda_{h}} $\cup$
$\bigcup_{h\in{}N}$\maximalparKomegaPlus{\lambda_{h}} =}
\end{displaymath}
\begin{displaymath}
\textrm{$\bigoplus_{h\in{}N}$\maximal{\eta_{h}} $\cup$
$\bigcup_{h\in{}N}$\maximalInf{\eta_{h}} $=$ \maximalInf{\sigma}
$=$ $\overline{K}^{\omega}$.}
\end{displaymath}
This concludes the proof.

\subsection{Proof of Lemma \ref{Lemma:From-MPAR-To-M-Inf}}

In order to prove Lemma \ref{Lemma:From-MPAR-To-M-Inf}, we need
the following Lemma.
\begin{Lemma}\label{Lemma:From-MKPAR-To-M} Let $p$
\multiderparK{\sigma} $p'$$\parallel$$p''$ with
$p,p',p''\in{}T_{PAR}$, $p'$ not containing occurrences of
$\hat{Z}_{F}$ and $\hat{Z}_{\infty}$, and $p''$ not containing
occurrences of variables in $Var$. Then, there exists a $t\in{}T$
such that $p$ \multider{\rho} $p'$$\parallel$$t$ with
\maximal{\rho} = \maximalparK{\sigma}, and $|\rho|>0$ if
$|\sigma|>0$.
\end{Lemma}
\begin{proof}
Let $\Re^{K}_{PAR}\setminus\Re=\{r_1,\ldots,r_m\}$, where for all
$i=1,\ldots,m$ $r_i$ is the $i$-th rule added into $\Re^{K}_{PAR}$
during the computation of algorithm of Lemma \ref{Lemma:Algo}. For
all $i=1,\ldots,m$ let us denote by $M^{K,i}_{PAR}$ (with support
$\Re^{K,i}_{PAR}$) the parallel \MBRS $M^{K}_{PAR}$ \emph{soon
before} the rule $r_i$ is added during the computation. Then, it
suffices to prove that the following two properties are satisfied:
\begin{enumerate}
    \item Let $p$
$\multidernorm{\sigma}{\Re^{K,i}_{PAR}}$ $p'$$\parallel$$p''$ with
$p,p',p''\in{}T_{PAR}$, $p'$ not containing occurrences of
$\hat{Z}_{F}$ and $\hat{Z}_{\infty}$, and $p''$ not containing
occurrences of variables in $Var$. Then, there exists a $t\in{}T$
such that $p$ \multider{\rho} $p'$$\parallel$$t$ with
\maximal{\rho} = \maximalG{M^{K,i}_{PAR}}{\sigma}, and $|\rho|>0$
if $|\sigma|>0$.
    \item If $r_i=X$\Rule{K'}$Y$ with $Y\in{}Var$ (resp.,
    $Y=\hat{Z}_{F}$), then there exists a derivation of the form
     $X$ \multider{\eta} $Y$ (resp., $X$ \multider{\eta} $t$ for some term
     $t$) such that \maximal{\eta} = $K'$ and $|\eta|>0$.
\end{enumerate}
The proof is by induction on $i$ (for the base step it is suffices
observe that $M^{K,1}_{PAR}=M_P$ where $M_P$ is the restriction of
$M$ to the PAR rules). For the inductive step Property 1 can be
easily proved  by induction on $\sigma$, while Property 2 follows
immediately by Property 1 and algorithm of Lemma \ref{Lemma:Algo}.
\end{proof}

\setcounter{equation}{0} In order to prove lemma
\ref{Lemma:From-MPAR-To-M-Inf} we use  a mapping for coding pairs
of  integers by single  integers. In particular, we consider the
following bijective mapping from $N\times{}N$ to $N$
\cite{davis83}
\begin{displaymath} \textrm{$<$ $>$$:
(x,y)\in{}N\times{}N\rightarrow{}2^{x}(2y+1)-1$}
\end{displaymath}
Let $\ell$ (resp. $\wp$) be the first (resp., second) component
 of $<$ $>^{-1}$.  Then,
\begin{enumerate}
    \item for all $x,y\in{}N$ $\ell(<$$x,y$$>)=x$ and
    $\wp(<$$x,y$$>)=y$,
    \item for all $z\in{}N$ $<$$\ell(z),\wp(z)$$>$ $=z$,
    \item for all $z\in{}N$ $\ell(z),\wp(z)\leq{}z$,
    \item for all $z,z'\in{}N$  if $z>z'$ and $\ell(z)=\ell(z')$ then $\wp(z)>\wp(z')$.
\end{enumerate}
Now, we introduce a new function
$next:N\times{}N\rightarrow{}N\times{}N$ defined as
\begin{displaymath}
 next(x,0)=(x,0)
\end{displaymath}
\begin{displaymath}
 next(x,y+1)= \left\{ \begin{array}{ll}
             (\ell(y),\wp(y)+1) & \textrm{if $next(x,y)=(\ell(y),\wp(y))$}\\
             next(x,y) & \textrm{otherwise}
             \end{array}
             \right.
\end{displaymath}

For all $x,y\in{}N$ let us denote by $next_{x}(y)$ the second
component of $next(x,y)$. The following lemma establishes some
properties of $next$. The proof is simple.

\begin{Lemma}\label{Lemma:next}
The function $next$ satisfies the following properties:
\begin{enumerate}
    \item For all $x,y\in{}N$ if $y\leq{}x$ then
              $next(x,y)=(x,0)$.
    \item For all $x,y\in{}N$ $next(x,y)=(x,z_{x,y})$ for some
    $z_{x,y}\in{}N$.
    \item  For all $x,y\in{}N$
          $next_{x}(y)\leq{}next_{x}(y+1)$.
    \item Let $x,y_{1},y_{2}\in{}N$ with
    $next_{x}(y_{1})<next_{x}(y_{2})$. Then, there exists a
    $k\in{}N$ such that $next(x,k)=(\ell(k),\wp(k))$, $\wp(k)=next_{x}(y_{2})-1$
    and $y_{1}\leq{}k<y_{2}$.
    \item For all $x,n\in{}N$  there exists a $y\in{}N$
    such that $next(x,y)=(x,n)$.
    \item For all $x\in{}N$  $next(\ell(x),x)=(\ell(x),\wp(x))$.
     \item For all  $x,i\in{}N$ if $i\neq\ell(x)$  then $next(i,x+1)=next(i,x)$.
\end{enumerate}
\end{Lemma}

\setcounter{equation}{0}Now, we can prove Lemma
\ref{Lemma:From-MPAR-To-M-Inf}. Let $p$ \multiderparKomega{\sigma}
with $p\in{}T_{PAR}$. We have to prove that there exists in $\Re$
a derivation of the form $p$ \multider{\delta} such that
\maximal{\delta} = \maximalparKomega{\sigma} and
\maximalInf{\delta} = \maximalparKomegaInf{\sigma} $\cup$
\maximalparKomegaPlus{\sigma}. Moreover, if $\sigma$ is either
infinite or contains some occurrence of rule in
$\Re^{K,K^{\omega}}_{PAR}\setminus\Re_{PAR}^{K}$, then $\delta$ is
infinite.
\newline Let $\lambda$ be the subsequence of
$\sigma$ containing all, and only, the occurrences of rules in
$\Re^{K,K^{\omega}}_{PAR}\setminus\Re_{PAR}^{K}$. Let us assume
that $\lambda$ is infinite. The proof for $\lambda$  finite (and
possibly empty) is simpler. Now, $\lambda=r_{0}r_{1}r_{2}\ldots$,
where for all $h\in{}N$
$r_{h}\in\Re^{K,K^{\omega}}_{PAR}\setminus\Re_{PAR}^{K}$.
Moreover, $\sigma$ can be written in the form
$\rho_{0}r_{0}\rho_{1}r_{1}\rho_{2}r_{2}\ldots$, where
$\sigma\setminus\lambda=\rho_{0}\rho_{1}\rho_{2}\ldots$ and for
all $h\in{}N$ $\rho_{h}$ is a finite rule sequence (possibly
empty) in $\Re_{PAR}^{K}$. For all $h\in{}N$ we denote by
$\sigma^{h}$ the suffix of $\sigma$ given by
$\rho_{h}r_{h}\rho_{h+1}r_{h+1}\ldots$.
 Now,
we prove  that there exists a sequence of terms in $T_{PAR}$,
 $(p_{h})_{h\in{N}}$, a sequence of variables
$(X_{h})_{h\in{}N}$ and a sequence of terms $(t_{h})_{h\in{}N}$
such that for all $h\in{}N$:
\begin{description}
    \item[i.] $p_{0}=p$,
    \item[ii.] $p_{h}$ \multiderparKomega{\sigma^{h}},
    \item[iii.] $p_{h}$ \multider{\eta_{h}}
    $p_{h+1}$$\parallel$$t_{h}$$\parallel$$X_{h}$ with \maximal{\eta_{h}} =
     \maximalparKomega{\rho_{h}},
    \item[iv.] $X_{h}$ \multider{\pi_{h}}
     with $\pi_{h}$ infinite, \maximal{\pi_{h}} =
     \maximalparKomega{r_{h}} and \maximalInf{\pi_{h}} =
    \maximalparKomegaPlus{r_{h}}.
\end{description}

Setting $p_{0}=p$, property ii is satisfied for $h=0$. So, let us
assume that the statement is true for all $h=0,\ldots,k$. Then, it
suffices to prove that
\begin{description}
    \item[A.] there exists a $p_{k+1}\in{}T_{PAR}$, a term $t_{k}$ and a variable
    $X_{k}$ such that $p_{k}$ \multider{\eta_{k}}
    $p_{k+1}$$\parallel$$t_{k}$$\parallel$$X_{k}$,   $p_{k+1}$ \multiderparKomega{\sigma^{k+1}},
    and $X_{k}$ \multider{\pi_{k}}
     with $\pi_{k}$ infinite. Moreover, \maximal{\eta_{k}} =
     \maximalparKomega{\rho_{k}}, \maximal{\pi_{k}} =
     \maximalparKomega{r_{k}} and \maximalInf{\pi_{k}} =
    \maximalparKomegaPlus{r_{k}}.
\end{description}
By the inductive hypothesis we have $p_{k}$
\multiderparKomega{\sigma^{k}}, that  can be written as
\begin{displaymath}
\textrm{$p_{k}$ \multiderparKomega{\rho_{k}}
$p'$$\parallel$$p''$$\parallel$$X$
        \onederparKomega{r_{k}} $p'$$\parallel$$p''$$\parallel$$\hat{Z}_{\infty}$
     \multiderparKomega{\sigma^{k+1}}}
\end{displaymath}
where $r_{k}=X$\Rule{K',K'^{\omega}}$\hat{Z}_{\infty}$ with
$X\in{}Var$ and $K',K'^{\omega}\in{}P_{n}$. Moreover, $p'$ does
not contain occurrences of $\hat{Z}_{F}$ and $\hat{Z}_{\infty}$,
and $p''$ doesn't contain occurrences of variables in $Var$. By
the definition of $\Re^{K,K^{\omega}}_{PAR}$ we have $X$
\multider{\pi_{k}}
     with $\pi_{k}$ infinite, \maximal{\pi_{k}} $=K'$ and \maximalInf{\pi_{k}}
     $=K'^{\omega}$. By Remark \ref{Remark:MPAR-Inf} we have
     \maximalparKomega{r_{k}} $=K'$ and
     \maximalparKomegaPlus{r_{k}} $=K'^{\omega}$.
Since the left-hand side of each rule in
$\Re^{K,K^{\omega}}_{PAR}$ does not contain occurrences of
$\hat{Z}_{F}$ and $\hat{Z}_{\infty}$, it follows that $p'$
\multiderparK{\sigma^{k+1}}. Since $\rho_{k}$ is a rule sequence
in $\Re_{PAR}^{K}$, by Lemma \ref{Lemma:From-MKPAR-To-M} it
follows that $p_{k}$ \multider{\eta_{k}}
    $p'$$\parallel$$t$$\parallel$$X$ for some term $t$ and
    \maximal{\eta_{k}} = \maximalparK{\rho_{k}}. By Remark \ref{Remark:MPAR-Inf}
    we deduce that \maximal{\eta_{k}} =
    \maximalparKomega{\rho_{k}}.
So, property \textbf{A} follows, setting $p_{k+1}=p'$,
$t_{k}=t$ and $X_{k}=X$. Thus, Properties i-iv are satisfied.\\
For all $h\in{}N$ the infinite derivation $X_{h}$
\multider{\pi_{h}} can be written as
\begin{equation}
\textrm{$s_{(h,0)}$ \prsoneder{r_{(h,0)}} $s_{(h,1)}$
\prsoneder{r_{(h,1)}} $s_{(h,2)}\ldots$}
\end{equation}
where $s_{(h,0)}=X_{h}$ and for all $k\in{}N$ $r_{(h,k)}\in\Re$.
For all $k,h\in{}N$ we denote by $\overline{r}_{k}$ the rule
$r_{(\ell(k),\wp(k))}$, and by $s_{h}(k)$ the term
$s_{next(h,k)}$. Now, we show that for all $k\in{}N$
\begin{eqnarray}
\textrm{$p_{k+1}$$\parallel$$t_{0}$$\parallel$$\ldots$$\parallel$$t_{k}$$\parallel$$s_{0}(k)$$\parallel$$s_{1}(k)$$\parallel$$\ldots$$\parallel$$s_{k}(k)$
        \multider{\eta_{k+1}\overline{r}_{k}}}\nonumber\\
\textrm{
$p_{k+2}$$\parallel$$t_{0}$$\parallel$$\ldots$$\parallel$$t_{k}$$\parallel$$t_{k+1}$$\parallel$$s_{0}(k+1)$$\parallel$$s_{1}(k+1)$$\parallel$$\ldots$$\parallel$$s_{k+1}(k+1)$
        }
\end{eqnarray}
By Lemma \ref{Lemma:next} it follows that
$s_{k}(k)=s_{next(k,k)}=s_{(k,0)}=X_{k}$. So, by Property iii we
deduce that
\begin{eqnarray}
\textrm{$p_{k+1}$$\parallel$$t_{0}$$\parallel$$\ldots$$\parallel$$t_{k}$$\parallel$$s_{0}(k)$$\parallel$$s_{1}(k)$$\parallel$$\ldots$$\parallel$$s_{k}(k)$
        \multider{\eta_{k+1}}}\nonumber\\
\textrm{
$p_{k+2}$$\parallel$$t_{0}$$\parallel$$\ldots$$\parallel$$t_{k}$$\parallel$$t_{k+1}$$\parallel$$s_{0}(k)$$\parallel$$s_{1}(k)$$\parallel$$\ldots$$\parallel$$s_{k}(k)$$\parallel$$s_{k+1}(k+1)$
        }
\end{eqnarray}
So, in order to obtain (2) it suffices to prove that
\begin{equation}
\textrm{$s_{0}(k)$$\parallel$$s_{1}(k)$$\parallel$$\ldots$$\parallel$$s_{k}(k)$
        \prsoneder{\overline{r}_{k}}
        $s_{0}(k+1)$$\parallel$$s_{1}(k+1)$$\parallel$$\ldots$$\parallel$$s_{k}(k+1)$
        }
\end{equation}
By Property 6 of Lemma \ref{Lemma:next} for all $k\in{}N$
$next(\ell(k),k)=(\ell(k),\wp(k))$. Moreover,
$next(\ell(k),k+1)=(\ell(k),\wp(k)+1)$. Therefore,
$s_{\ell(k)}(k)=s_{(\ell(k),\wp(k))}$ \prsoneder{\overline{r}_{k}}
$s_{(\ell(k),\wp(k)+1)}=s_{\ell(k)}(k+1)$. By Property 7 of Lemma
\ref{Lemma:next} for all $i\neq{}\ell(k)$ $next(i,k+1)=next(i,k)$.
So, for all $i\neq{}\ell(k)$ $s_{i}(k+1)=s_{i}(k)$. Since
$\ell(k)\leq{}k$,  we obtain evidently (4). So,  (2) is satisfied
for all $k\in{}N$. Moreover, since $s_{0}(0)=X_{0}$, we have
\begin{equation}
\textrm{$p=p_{0}$ \multider{\eta_{0}}
        $p_{1}$$\parallel$$t_{0}$$\parallel$$s_{0}(0)$
        }
\end{equation}
Setting
$\delta=\eta_{0}\eta_{1}\overline{r}_{0}\eta_{2}\overline{r}_{1}\eta_{3}\overline{r}_{2}\ldots$,
from (2) and (5) we obtain that $p$ \multider{\delta} with
$\delta$ infinite. Therefore, it remains to prove that
\maximal{\delta} = \maximalparKomega{\sigma} and
\maximalInf{\delta} = \maximalparKomegaInf{\sigma} $\cup$
\maximalparKomegaPlus{\sigma}. Let
$\mu=\overline{r}_{0}\overline{r}_{1}\overline{r}_{2}\ldots$.
Evidently, $\mu\in{}Interleaving((\pi_{h})_{h\in{}N})$.
 By Properties iii-iv, Proposition \ref{Prop:MaximalInterleaving},
 and remembering that
$\sigma=\rho_{0}r_{0}\rho_{1}r_{1}\ldots$, we obtain
\begin{displaymath}
\textrm{\maximal{\delta} = $\bigcup_{h\in{}N}$\maximal{\eta_{h}}
$\cup$ \maximal{\mu} =
$\bigcup_{h\in{}N}$\maximalparKomega{\rho_{h}} $\cup$
$\bigcup_{h\in{}N}$\maximal{\pi_{h}}
        =}
\end{displaymath}
\begin{displaymath}
\textrm{$\bigcup_{h\in{}N}$\maximalparKomega{\rho_{h}} $\cup$
$\bigcup_{h\in{}N}$\maximalparKomega{r_{h}} =
\maximalparKomega{\sigma}.}
\end{displaymath}
By Remark \ref{Remark:MPAR-Inf}, for all $r\in\Re_{PAR}^{K}$
\maximalparKomegaPlus{r} = $\emptyset$. Remembering that
$\lambda=r_{0}r_{1}r_{2}\ldots$, by Properties iii-iv and
Proposition  \ref{Prop:MaximalInterleaving} we obtain
\begin{displaymath}
\textrm{\maximalInf{\delta} =
$\bigoplus_{h\in{}N}$\maximal{\eta_{h}} $\cup$ \maximalInf{\mu} =
$\bigoplus_{h\in{}N}$\maximalparKomega{\rho_{h}} $\cup$
$\bigcup_{h\in{}N}$\maximalInf{\pi_{h}} $\cup$
$\bigoplus_{h\in{}N}$\maximal{\pi_{h}}
        =}
\end{displaymath}
\begin{displaymath}
\textrm{\maximalparKomegaInf{\sigma\setminus\lambda} $\cup$
$\bigcup_{h\in{}N}$\maximalparKomegaPlus{r_{h}} $\cup$
$\bigoplus_{h\in{}N}$\maximalparKomega{r_{h}} = }
\end{displaymath}
\begin{displaymath}
\textrm{\maximalparKomegaInf{\sigma\setminus\lambda} $\cup$
\maximalparKomegaPlus{\sigma} $\cup$ \maximalparKomegaInf{\lambda}
= \maximalparKomegaInf{\sigma} $\cup$
\maximalparKomegaPlus{\sigma}.}
\end{displaymath}
This concludes the proof.

\section{Proof of Theorems \ref{Theorem:Problem1.1} and \ref{Theorem:Problem1.2}}

In order to prove Theorems \ref{Theorem:Problem1.1} and
\ref{Theorem:Problem1.2} we need the following lemmata
\ref{Lemma:From-MKSEQ-To-M}--\ref{Lemma:From-M-To-MSEQ3}.

\begin{Remark}\label{Remark:MKSEQ} By construction the following
properties hold
\begin{itemize}
    \item  for all $r\in\Re\cap\Re_{SEQ}^{K}\quad$
        \maximal{r} = \maximalseqK{r}.
    \item for all $r=X$\Rule{K'}$Y\in\Re_{SEQ}^{K}\setminus\Re\quad$  \maximalseqK{r}
            = $K'$.
\end{itemize}
\end{Remark}

\setcounter{equation}{0}
\begin{Lemma}\label{Lemma:From-MKSEQ-To-M}
Let $t,t'\in{}T_{SEQ}$ and $s$ be any term in $T$ such that
$t\in{}SEQ(s)$. The following results hold
\begin{enumerate}
    \item  If $t$ \onederseqK{r} $t'$ with $r\in\Re^{K}_{SEQ}$, then
     there exists a $s'\in{}T$ with $t'\in{}SEQ(s')$ such that $s$ \multider{\sigma} $s'$,
     with \maximal{\sigma} = \maximalseqK{r} and $|\sigma|>0$.
      \item If $t$ \multiderseqK{\sigma} $t'$ with $t\neq\varepsilon$,
    then there exists a $s'\in{}T$ with $t'\in{}SEQ(s')$ such that $s$
    \multider{\rho} $s'$, with \maximal{\rho} =
    \maximalseqK{\sigma}, and $|\rho|>0$ if $|\sigma|>0$.
    \item If $t$ \multiderseqK{\sigma} is a $(K,K^{\omega})$-accepting infinite derivation  in
    $M_{SEQ}^{K}$ from $t\in{}T_{SEQ}$, then  there exists a $(K,K^{\omega})$-accepting infinite derivation
     in $M$ from s.
\end{enumerate}
\end{Lemma}
\begin{proof}
At first, we prove Property 1. There are two cases:
\begin{itemize}
    \item  $r=Y$\Rule{a}$Z_{1}.Z_{2}\in{}\Re$.  By Remark \ref{Remark:MKSEQ}
     \maximal{r} = \maximalseqK{r}. Since  $t\in{}SEQ(s)$ and $t$ \onederseqK{r} $t'$,
     we deduce that
     there exists a $s'\in{}s[Y\rightarrow{}Z_{1}.Z_{2}]$
      such that $t'\in{}SEQ(s')$. Since $Y$ \prsoneder{r}
      $Z_{1}.Z_{2}$,
      by Proposition \ref{Prop:Subterms1} it follows that $s$ \prsoneder{r} $s'$.
      Therefore, Property 1 is satisfied.
    \item $r=Y$\Rule{K'}$Z$ with $Y,Z\in{}Var$,
    \maximalseqK{r} = $K'$, $last(t)=Y$ and $last(t')=Z$.
     By the definition of $\Re^{K}_{SEQ}$ there exists a derivation
      in
     $\Re^{K}_{PAR}$ of the form $Y$ \multiderparK{\sigma} $p$$\parallel$$Z$
     for some
     $p\in{}T_{PAR}$, with \maximalparK{\sigma} = \maximalseqK{r} and
     $|\sigma|>0$.
      By Lemma \ref{Lemma:From-MKPAR-To-M} there exists a term
      $st$ such that $Y$ \multider{\rho} $st$$\parallel$$Z$ with
      \maximal{\rho} = \maximalparK{\sigma} and $|\rho|>0$. So, \maximal{\rho} = \maximalseqK{r}.
       Since  $t\in{}SEQ(s)$ and $t$ \onederseqK{r} $t'$, we deduce that there exists a
    $s'\in{}s[Y\rightarrow{}st$$\parallel$$Z]$
    such that
    $t'\in{}SEQ(s')$.
    Since $Y$ \multider{\rho} $st$$\parallel$$Z$,
    by Proposition \ref{Prop:Subterms1} we conclude that $s$ \multider{\rho} $s'$
    with $|\rho|>0$. Thus, Property 1 is satisfied.
\end{itemize}
Property 2 can be easily proved by induction on the length of
$\sigma$, and using  Property 1. Finally, Property 3 easily
follows  from Property 1 and Proposition
\ref{Prop:MaximalInterleaving}.
\end{proof}

\noindent The following definition introduces the notion of level
of application of a rule in a derivation:

\begin{Definition}
  Let $t$ \prsoneder{r} $t'$  be a single--step derivation in $\Re$ with $t\in{}T$.
  We say that $r$ is \emph{applicable at level 0} in
  $t$ \prsoneder{r} $t'$, if $t = \overline{t}$$\parallel$$s$,
  $t' = \overline{t}$$\parallel$$s'$ $\mathrm{(}$for some $\overline{t}, s, s'\in
  T$$\mathbb{)}$, and $r = s$\Rule{a}$s'$, for some $a\in\Sigma$.

  We say that $r$ is \emph{applicable at level $k > 0$ in}
  $t$ \prsoneder{r} $t'$, if $t = \overline{t}$$\parallel$$(X.s)$,
  $t' = \overline{t}$$\parallel$$(X.s')$ $\mathrm{(}$for some
  $\overline{t}, s,s'\in T$$\mathbb{)}$, $s$ \prsoneder{r} $s'$, and
  $r$ is applicable at level $k-1$ in $s$ \prsoneder{r} $s'$.

  The \emph{level of application}  of $r$ in $t$ \prsoneder{r}
  $t'$ is the greatest level of applicability of $r$ in $t$ \prsoneder{r}
  $t'$.
\end{Definition}

The definition above extends in the obvious way to $n$--step
derivations and to infinite derivations.

\begin{Lemma}\label{Lemma:From-M-To-MSEQ2}
 Let $i\in{}K$, $X\in{}Var$ and
 $X$ \multider{\sigma} be a
 $(K,K^{\omega})$-accepting infinite derivation  in $M$ from
 $X$.
Then, one of the following conditions is satisfied:
\begin{enumerate}
    \item There exists a variable  $Y\in{}Var$ reachable from $X$ in
    $\Re_{SEQ}^{K}$ through a $(K',\emptyset)$-accepting derivation in $M_{SEQ}^{K}$
    with $K'\subseteq{}K$, and there exists a derivation
    $Y$ \multiderparKomega{\rho} such that \maximalparKomega{\rho} = $K$
    and
    \maximalparKomegaInf{\rho} $\cup$ \maximalparKomegaPlus{\rho} =
    $K^{\omega}$. Moreover,  $\rho$ is either
    infinite or  contains some occurrence of rule in
    $\Re^{K,K^{\omega}}_{PAR}\setminus\Re_{PAR}^{K}$.
    \item There exists a variable  $Y\in{}Var$ reachable from $X$ in
    $\Re_{SEQ}^{K}$ through a $(K_{i},\emptyset)$-accepting derivation in $M_{SEQ}^{K}$
    with $\{i\}\subseteq{}K_{i}\subseteq{}K$, and
    there exists a $(K,K^{\omega})$-accepting infinite derivation  in $M$ from
    $Y$.
\end{enumerate}
\end{Lemma}
\begin{proof}
The proof is by induction on the level $k$ of application of the
first occurrence of a rule $r$ of $\Re_{i}^{A}$  in a
$(K,K^{\omega})$-accepting infinite derivation  in $M$ from
 a variable. If $X$ \multider{\sigma} is in
$\Pi^{K,K^{\omega}}_{PAR,\infty}$, by Lemma
\ref{Lemma:From-M-To-MPAROMEGA} Property 1 follows, setting $Y=X$.
Otherwise,  it is easy to deduce  that the derivation $X$
\multider{\sigma} can be written in the form
\begin{displaymath}
\textrm{$X$ \multider{\sigma_{1}} $t$$\parallel$$Z$ \prsoneder{r'}
$t$$\parallel$$W.Z'$ \multider{\sigma_{2}}}
\end{displaymath}
where $r'=Z$\Rule{a}$W.Z'$ (with $W,Z,Z'\in{}Var$), and there
exists a subderivation of $t$$\parallel$$W.Z'$
\multider{\sigma_{2}} from $Z'$, namely $Z'$
\multider{\sigma_{2}'}, that is a $(K,K^{\omega})$-accepting
infinite derivation in $M$.\\
\textbf{Base Step}: $k=0$.  In this case $r$ must occur in the
rule sequence $\sigma_{1}r'(\sigma_{2}\setminus\sigma_{2}')$.
 By Lemma \ref{Lemma:Subderivations1}, we have $t$
\multider{\sigma_{2}\setminus\sigma_{2}'}. Therefore,  there
exists a derivation of the form $X$ \multider{\lambda}
$t'$$\parallel$$Z$ \prsoneder{r'} $t'$$\parallel$$W.Z'$ with
$\{i\}\subseteq$ \maximal{\lambda{}r'} $\subseteq{}K$. By Lemma
 \ref{Lemma:From-M-To-MKPAR}, applied to the derivation $X$ \multider{\lambda}
$t'$$\parallel$$Z$, there exists a $p\in{}T_{PAR}$ such that $X$
\multiderparK{\rho} $p$$\parallel$$Z$, with \maximalparK{\rho} =
\maximal{\lambda}.  By the definition of $\Re_{SEQ}^{K}$ we have
 that $X$ \multiderseqK{\mu} $Z$ \onederseqK{r'}
$W.Z'$, with \maximalseqK{\mu} = \maximalparK{\rho} and
\maximalseqK{r'} = \maximal{r'}. Therefore, \maximalseqK{\mu{}r'}
= \maximal{\lambda{}r'}.
 Thus,  variable $Z'$ is reachable from $X$ in
    $\Re_{SEQ}^{K}$ through a $(K_{i},\emptyset)$-accepting derivation in $M_{SEQ}^{K}$
    with $\{i\}\subseteq{}K_{i}\subseteq{}K$, and
    there exists a $(K,K^{\omega})$-accepting infinite derivation  in $M$ from
    $Z'$. This is exactly what Property 2 states.\\
 \textbf{Induction Step}: $k>0$. If the rule sequence
    $\sigma_{1}r'( \sigma_{2}\setminus\sigma_{2}')$ contains some occurrence of $r$,
    then the thesis follows by reasoning as
    in the base step. Otherwise, $\sigma_{2}'$ contains the first occurrence of $r$ in $\sigma$.
    Clearly, this occurrence is the first occurrence of a  rule of $\Re_{i}^{A}$
    in the $(K,K^{\omega})$-accepting
    infinite derivation
     $Z'$ \multider{\sigma_{2}'}, and it is applied at level $k'$ in
    $Z'$ \multider{\sigma_{2}'} with $k'<k$. By inductive hypothesis, the thesis holds for
    the derivation $Z'$ \multider{\sigma_{2}'}.
    Therefore, it suffices to prove that  $Z'$
    is reachable from  $X$ in $\Re_{SEQ}^{K}$
    through a $(K',\emptyset)$-accepting derivation in $M_{SEQ}^{K}$
    with $K'\subseteq{}K$. By Lemma \ref{Lemma:From-M-To-MKPAR}, applied
    to the
    derivation   $X$ \multider{\sigma_{1}} $t$$\parallel$$Z$, there exists a $p\in{}T_{PAR}$
    such that $X$ \multiderparK{\rho} $p$$\parallel$$Z$
    with \maximalparK{\rho} = \maximal{\sigma_{1}} $\subseteq{}K$. By the
    definition of $\Re_{SEQ}^{K}$ we obtain that  $X$
    \multiderseqK{\mu} $Z$ \onederseqK{r'} $W.Z'$
    with  \maximalseqK{\mu} = \maximalparK{\rho} and
    \maximalseqK{r'} = \maximal{r'} $\subseteq{}K$.
    So, \maximalseqK{\mu{}r'} $\subseteq{}K$. This concludes the proof.
\end{proof}

\begin{Lemma}\label{Lemma:From-M-To-MSEQ3}
 Let  $X\in{}Var$ and
 $X$ \multider{\sigma} be a
 $(K,K^{\omega})$-accepting infinite derivation  in $M$ from
 $X$.
Then, one of the following conditions is satisfied:
\begin{enumerate}
    \item There exists a variable  $Y\in{}Var$ reachable from $X$ in
    $\Re_{SEQ}^{K}$ through a $(K',\emptyset)$-accepting derivation in $M_{SEQ}^{K}$
    with $K'\subseteq{}K$, and there exists a derivation
    $Y$ \multiderparKomega{\rho} such that \maximalparKomega{\rho} = $K$
    and
    \maximalparKomegaInf{\rho} $\cup$ \maximalparKomegaPlus{\rho} =
    $K^{\omega}$.
    Moreover,  $\rho$ is either
    infinite or  contains some occurrence of rule in
    $\Re^{K,K^{\omega}}_{PAR}\setminus\Re_{PAR}^{K}$.
    \item There exists a variable  $Y\in{}Var$ reachable from $X$ in
    $\Re_{SEQ}^{K}$ through a $(K,\emptyset)$-accepting derivation in $M_{SEQ}^{K}$, and
    there exists a $(K,K^{\omega})$-accepting infinite derivation  in $M$ from
    $Y$.
\end{enumerate}
\end{Lemma}
\begin{proof}
It suffices to prove that, assuming that Property 1 is not
satisfied, Property 2 must hold. If $|K|=0$, property 2  is
obviously satisfied. So, let us assume that $|K|>0$. Let
$K=\{j_{1},\ldots,j_{|K|}\}$, and for all $p=1,\ldots,|K|$ let
$K_{p}=\{j_{1},\ldots,j_{p}\}$. Let us prove by induction on $p$
that  the following property is satisfied :
\begin{description}
    \item[A] There exists a variable  $Y$ reachable from $X$ in
    $\Re_{SEQ}^{K}$ through a $(K',\emptyset)$-accepting derivation in $M_{SEQ}^{K}$
    with $K_{p}\subseteq{}K'\subseteq{}K$, and
    there exists a $(K,K^{\omega})$-accepting infinite derivation  in $M$ from
    $Y$.
\end{description}
 \textbf{Base Step}: $p=1$. Considering that Property 1 isn't satisfied,
 the result follows from Lemma \ref{Lemma:From-M-To-MSEQ2},
 setting $i=j_{1}$.\\
 \textbf{Induction Step}: $1<p\leq{}|K|$.
By the inductive hypothesis there exists a
$t\in{}T_{SEQ}\setminus\{\varepsilon\}$ such that $X$
\multiderseqK{\rho} $t$ with $K_{p-1}\subseteq{}$
\maximalseqK{\rho} $\subseteq{}K$, and there exists a
$(K,K^{\omega})$-accepting infinite derivation in $M$ of the form
$last(t)$ \multider{\eta}. By Lemma \ref{Lemma:From-M-To-MSEQ2},
applied to the derivation $last(t)$ \multider{\eta}, and
considering that  Property 1 is not satisfied, it follows that
there exists a $\overline{t}\in{}T_{SEQ}\setminus\{\varepsilon\}$
such that $last(t)$ \multiderseqK{\overline{\rho}} $\overline{t}$
with $\{j_{p}\}\subseteq{}$\maximalseqK{\overline{\rho}}
$\subseteq{}K$, and there exists a $(K,K^{\omega})$-accepting
infinite derivation in $M$ from $last(\overline{t})$. So, we have
$X$ \multiderseqK{\rho\overline{\rho}} $t\circ{}\overline{t}$ with
$K_{p}\subseteq{}$\maximalseqK{\rho\overline{\rho}}
$\subseteq{}K$.
Therefore, Property \textbf{A} follows, setting $Y=last(\overline{t})$.\\
By property \textbf{A}, the thesis follows.
\end{proof}

\subsection{Proof of Theorem \ref{Theorem:Problem1.1}}

($\Rightarrow$) Since $K\neq{}K^{\omega}$ and
$K\supseteq{}K^{\omega}$, it follows that $K\supset{}K^{\omega}$.
 Let  $d=X$ \multider{\sigma} be
 a $(K,K^{\omega})$-accepting infinite derivation  in $M$ from
 $X$.  Evidently,
$K\setminus{}K^{\omega}=\{i\in\{1,\ldots,n\}|$ $\sigma$ contains a
finite non--null number of occurrences of rules in
$\Re_{i}^{A}\}$. Then, for all $i\in{}K\setminus{}K^{\omega}$ it
is defined the greatest application level, denoted by $h_{i}(d)$,
of occurrences of rules of $\Re_{i}^{A}$ in the derivation $d$.
The proof is by induction on
$max_{i\in{}K\setminus{}K^{\omega}}\{h_{i}(d)\}$.\\
 \textbf{Base Step}:
$max_{i\in{}K\setminus{}K^{\omega}}\{h_{i}(d)\}=0$. In this case
 each subderivation  of  $d=X$ \multider{\sigma}
does not contain occurrences of rules in
$\bigcup_{i\in{}K\setminus{}K^{\omega}}\Re_{i}^{A}$. So, $d$
belongs to
 $\Pi^{K,K^{\omega}}_{PAR,\infty}$. Then, by Lemma
 \ref{Lemma:From-M-To-MPAROMEGA} we obtain the assertion
 setting $Y=X$.\\
 \textbf{Induction Step}: $max_{i\in{}K\setminus{}K^{\omega}}\{h_{i}(d)\}>0$. If
 $d=X$ \multider{\sigma} is in
$\Pi^{K,K^{\omega}}_{PAR,\infty}$, by Lemma
\ref{Lemma:From-M-To-MPAROMEGA} we obtain the assertion setting
$Y=X$. Otherwise, it is easy to deduce that the derivation $X$
\multider{\sigma} can be written in the form
\begin{displaymath}
\textrm{$X$ \multider{\sigma_{1}} $t$$\parallel$$Z$ \prsoneder{r}
$t$$\parallel$$W.Z'$ \multider{\sigma_{2}}}
\end{displaymath}
where $r=Z$\Rule{a}$W.Z'$ (with $W,Z,Z'\in{}Var$), and there
exists a subderivation of $t$$\parallel$$W.Z'$
\multider{\sigma_{2}} from $Z'$, namely $d'=Z'$
\multider{\sigma_{2}'}, that  is a $(K,K^{\omega})$-accepting
infinite derivation in $M$. Evidently,
$max_{i\in{}K\setminus{}K^{\omega}}\{h_{i}(d')\}<max_{i\in{}K\setminus{}K^{\omega}}\{h_{i}(d)\}$.
By inductive hypothesis, the thesis holds for the derivation $d'$.
Therefore, it suffices to prove that $Z'$ is reachable from $X$ in
    $\Re_{SEQ}^{K}$ through a $(K',\emptyset)$-accepting derivation in $M_{SEQ}^{K}$
    with $K'\subseteq{}K$. By Lemma \ref{Lemma:From-M-To-MKPAR}, applied to
    the derivation    $X$ \multider{\sigma_{1}} $t$$\parallel$$Z$
    where \maximal{\sigma_{1}} $\subseteq{}K$, there exists a $p\in{}T_{PAR}$
    such that $X$ \multiderparK{\rho_{1}} $p$$\parallel$$Z$
    with \maximalparK{\rho_{1}} = \maximal{\sigma_{1}}. By the
    definition of $\Re_{SEQ}^{K}$ we obtain that  $X$
    \multiderseqK{\gamma} $Z$ \onederseqK{r} $W.Z'$,
    with \maximalseqK{\gamma} = \maximalparK{\rho_{1}} and
    \maximalseqK{r} = \maximal{r} $\subseteq{}K$. So,
    \maximalseqK{\gamma{}r} $\subseteq{}K$. Therefore, the thesis
    holds.\newline

\noindent($\mathbb{\Leftarrow}$) By the hypothesis we have
\begin{enumerate}
    \item $X$ \multiderseqK{\lambda} $t$ with
    $t\in{}T_{SEQ}\setminus\{\varepsilon\}$, $last(t)=Y$ and
    \maximalseqK{\lambda} $\subseteq{}K$.
    \item $Y$ \multiderparKomega{\rho}  with \maximalparKomega{\rho} = $K$
    and
    \maximalparKomegaInf{\rho} $\cup$ \maximalparKomegaPlus{\rho} =
    $K^{\omega}$. Moreover,  $\rho$ is either
    infinite or  contains some occurrence of rule in
    $\Re^{K,K^{\omega}}_{PAR}\setminus\Re_{PAR}^{K}$.
\end{enumerate}
Since $X\in{}SEQ(X)$, by condition 1 and Lemma
\ref{Lemma:From-MKSEQ-To-M}, it follows that there exists a
$s\in{}T$ such that $t\in{}SEQ(s)$ and $X$ \multider{\eta} $s$
with \maximal{\eta} $\subseteq{}K$. By condition 2 and  Lemma
\ref{Lemma:From-MPAR-To-M-Inf} it follows that there exists  a
 $(K,K^{\omega})$-accepting infinite derivation  in $M$
of the form $Y$ \multider{\sigma}. Since $Y\in{}SubTerms(s)$, by
Proposition \ref{Prop:Subterms1} we have that $s$
\multider{\sigma}. After all, we obtain $X$ \multider{\eta} $s$
\multider{\sigma}, that is  a
 $(K,K^{\omega})$-accepting infinite derivation  in $M$ from $X$.
 This concludes the proof.

\subsection{Proof of Theorem \ref{Theorem:Problem1.2}}

($\Rightarrow$) It suffices to prove that, assuming that condition
1 (in the enunciation) does not hold, condition 2 must hold. Under
this hypothesis, we show that there exists a sequence of terms
  $(t_{h})_{h\in{}N}$ in
$T_{SEQ}\setminus\{\varepsilon\}$ satisfying  the following
properties for all $h\in{}N$:
\begin{description}
    \item[i.] $t_{0}=X$,
    \item[ii.] $last(t_{h})$
    \multiderseqK{\rho_{h}} $t_{h+1}$ with \maximalseqK{\rho_{h}}
    $=K$,
    \item[iii.]  there exists a
    $(K,K^{\omega})$-accepting infinite derivation  in $M$ from
    $last(t_{h})$,
    \item[iv.] $last(t_{h})$ is
    reachable from $X$ in $\Re_{SEQ}^{K}$ through a $(K',\emptyset)$-accepting
    derivation in $M_{SEQ}^{K}$ with $K'\subseteq{}K$.
\end{description}
For $h=0$ properties iii and iv are satisfied,  by setting
$t_{0}=X$. So, assume the existence of a finite  sequence of terms
$t_{0},t_{1},\ldots,t_{h}$ in $T_{SEQ}\setminus\{\varepsilon\}$
sa\-ti\-sfying  properties i-iv. It suffices to prove that there
exists a term $t_{h+1}$ in $T_{SEQ}\setminus\{\varepsilon\}$
satisfying  iii and iv, and such that  $last(t_{h})$
    \multiderseqK{\rho_{h}} $t_{h+1}$ with \maximalseqK{\rho_{h}} $=K$.
By the inductive hypothesis, $last(t_{h})$ is reachable from
    $X$ in $\Re_{SEQ}^{K}$ through a $(K',\emptyset)$-accepting
    derivation in $M_{SEQ}^{K}$ with $K'\subseteq{}K$, and there exists a $(K,K^{\omega})$-accepting
    infinite derivation in $M$ from $last(t_{h})$.
    By Lemma \ref{Lemma:From-M-To-MSEQ3} applied to  variable $last(t_{h})$,
    and the fact that
     condition 1 does not hold, it follows that there exists a term
   $t\in{}T_{SEQ}\setminus\{\varepsilon\}$ such that
    $last(t_{h})$
    \multiderseqK{\rho_{h}} $t$ with \maximalseqK{\rho_{h}} $=K$, and
    there exists a $(K,K^{\omega})$-accepting infinite derivation in $M$ from
    $last(t)$. Since $last(t_{h})$ is reachable from
    $X$ in $\Re_{SEQ}^{K}$ through a $(K',\emptyset)$-accepting
    derivation in $M_{SEQ}^{K}$ with $K'\subseteq{}K$,  it follows that  $last(t)$ is reachable
    from
    $X$ in $\Re_{SEQ}^{K}$ through a $(K,\emptyset)$-accepting
    derivation in $M_{SEQ}^{K}$. Thus, setting
    $t_{h+1}=t$, we obtain the result.\\
    Let $(t_{h})_{h\in{}N}$ be the sequence of terms in  $T_{SEQ}\setminus\{\varepsilon\}$
     satisfying  properties i-iv. Since in this case $|K|>0$ (remember that $|K|+|K^{\omega}|>0$),
     we have $|\rho_{h}|>0$  for all $h\in{}N$. Then, by Proposition
     \ref{Prop:Subterms2} we obtain that for all
     $h\in{}N$
\begin{displaymath}
\textrm{
 $t_{0}$$\circ$$t_{1}$$\circ$$\ldots$$\circ$$t_{h}$
    \multiderseqK{\rho_{h}} $t_{0}$$\circ$$t_{1}$$\circ$$\ldots$$\circ$$t_{h}$$\circ$$t_{h+1}$
     }
\end{displaymath}
Therefore,
\begin{eqnarray}
\textrm{$X=t_{0}$ \multiderseqK{\rho_{0}} $t_{0}$$\circ$$t_{1}$
\multiderseqK{\rho_{1}} $t_{0}$$\circ$$t_{1}$$\circ$$t_{2}$
\multiderseqK{\rho_{2}} $\ldots$ \multiderseqK{\rho_{h-1}}
$t_{0}$$\circ$$t_{1}$$\circ$$\ldots$$\circ$$t_{h}$}\nonumber\\
\textrm{\multiderseqK{\rho_{h}}
$t_{0}$$\circ$$t_{1}$$\circ$$\ldots$$\circ$$t_{h}$$\circ$$t_{h+1}$
\multiderseqK{\rho_{h+1}}$\ldots$}\nonumber
\end{eqnarray}
 is an  infinite derivation in
$\Re_{SEQ}^{K}$ from $X$. Setting $\delta=\rho_{0}\rho_{1}\ldots$,
from ii and Proposition \ref{Prop:MaximalInterleaving} we obtain
that \maximalseqK{\delta} =
$\bigcup_{h\in{}N}$\maximalseqK{\rho_{h}} = $K$ and
\maximalseqInfK{\delta} =
$\bigoplus_{h\in{}N}$\maximalseqK{\rho_{h}} = $K$ = $K^{\omega}$.
Hence, condition 2 (in the enunciation) holds.\\

\noindent($\Leftarrow$) At first, let us assume that  condition 2
holds. Then, since $X\in{}SEQ(X)$, the result follows directly by
Lemma \ref{Lemma:From-MKSEQ-To-M}. Assume that condition 1 holds
instead. Then, we reason as in the proof of Theorem
\ref{Theorem:Problem1.1}.

\end{document}